\documentclass[preprint,journal, hideappendix]{vgtc}            

\onlineid{1671}

\vgtccategory{Research}

\vgtcpapertype{Analytics \& Decisions}

\title{CausalPrism: A Visual Analytics Approach for Subgroup-based Causal Heterogeneity Exploration}

\author{%
  \authororcid{Jiehui Zhou}{0000-0003-0709-775X},
  Xumeng Wang, Kam-Kwai Wong, Wei Zhang, Xingyu Liu, Juntian Zhang, Minfeng Zhu and 
  Wei Chen
}

\authorfooter{
  \item
  	J. Zhou, W. Zhang, X. Liu, J. Zhang, M. Zhu and W. Chen are with the State Key Lab of CAD\&CG, Zhejiang University.
  	E-mail: \{zhoujiehui, 22151190, liu\_xingyu, 3200105799, minfeng\_zhu, chenvis\}@zju.edu.cn.
  \item
  	X. Wang is with TMCC, CS, Nankai University.
  	E-mail: wangxumeng@nankai.edu.cn.

  \item KK. Wong is with Hong Kong University of Science and Technology and Georgia Institute of Technology.
  	E-mail: kkwongar@connect.ust.hk.
}

\abstract{%
  In causal inference, estimating Heterogeneous Treatment Effects (HTEs) from observational data is critical for understanding how different subgroups respond to treatments, with broad applications such as precision medicine and targeted advertising. However, existing work on HTE, subgroup discovery, and causal visualization is insufficient to address two challenges: first, the sheer number of potential subgroups and the necessity to balance multiple objectives (\eg high effects and low variances) pose a considerable analytical challenge. Second, effective subgroup analysis has to follow the analysis goal specified by users and provide causal results with verification. To this end, we propose a visual analytics approach for subgroup-based causal heterogeneity exploration. Specifically, we first formulate causal subgroup discovery as a constrained multi-objective optimization problem and adopt a heuristic genetic algorithm to learn the Pareto front of optimal subgroups described by interpretable rules. Combining with this model, we develop a prototype system, \sysname, that incorporates tabular visualization, multi-attribute rankings, and uncertainty plots to support users in interactively exploring and sorting subgroups and explaining treatment effects. Quantitative experiments validate that the proposed model can efficiently mine causal subgroups that outperform state-of-the-art HTE and subgroup discovery methods, and case studies and expert interviews demonstrate the effectiveness and usability of the system. Code is available at \href{https://osf.io/jaqmf/?view_only=ac9575209945476b955bf829c85196e9}{OSF}.
}


\teaser{
  \centering
  \includegraphics[width=\linewidth]{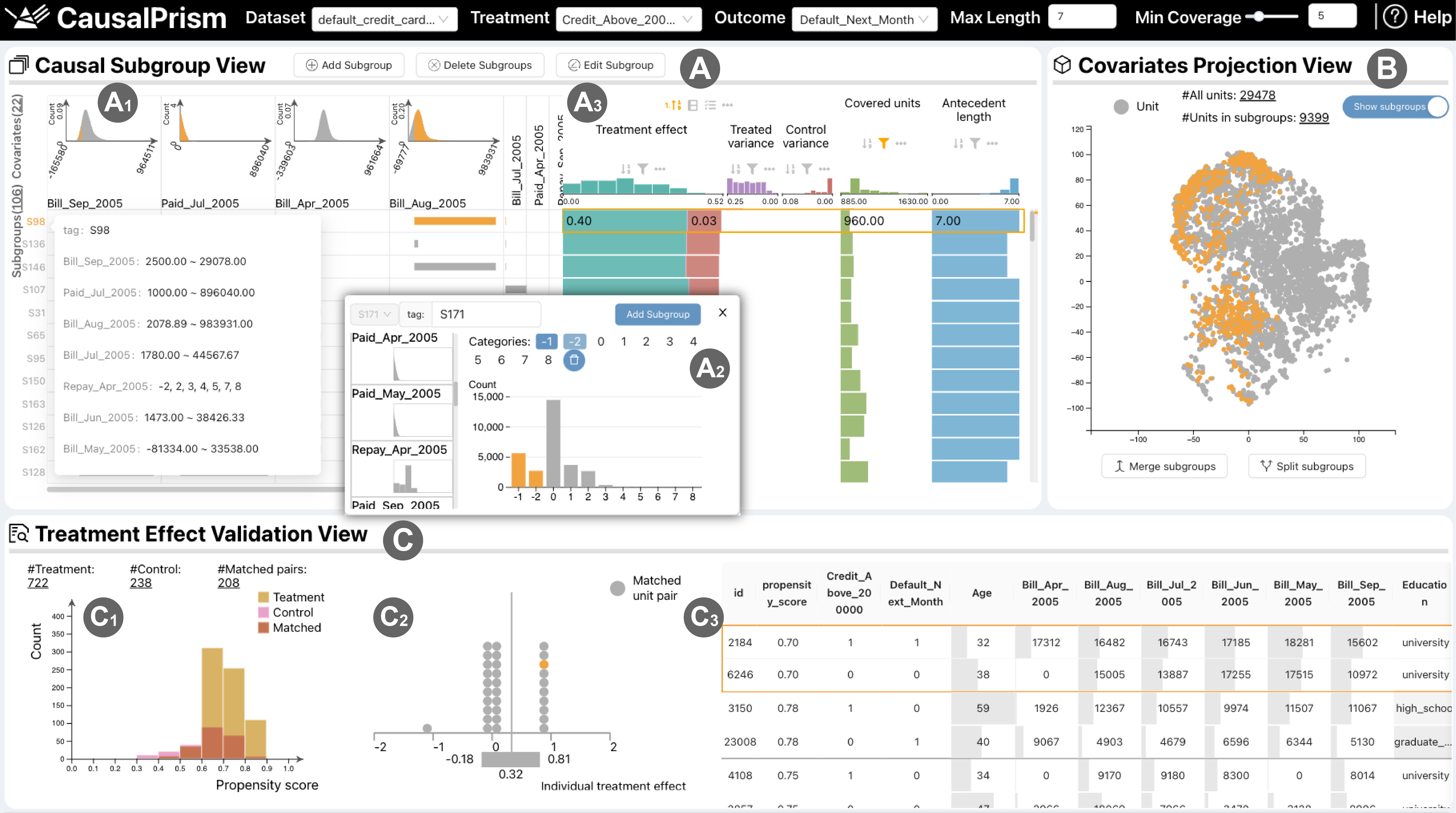}
  \caption{%
  	\sysname helps identify, explore, rank, and interprete causal subgroups in observational data. (A) The Causal Subgroup View includes a tailored tabular visualization of subgroup descriptions, a subgroup editing window, and a ranking visualization of multiple evaluation metrics to support subgroup overview, modification, and ranking. (B) The Covariate Projection View reduces units with high-dimensional covariates to two dimensions, allowing users to analyze similarities between subgroups and assisting in merging and splitting subgroups. (C) The Treatment Effect Validation View consists of propensity score histograms, individual treatment effect dot plots, and detailed information of matched units, which helps users interpret effect strength and uncertainty, thereby increasing trust.%
  }
  \label{fig:teaser}
}

\graphicspath{{figs/}{figures/}{pictures/}{images/}{./}} 

\usepackage{tabu}                      
\usepackage{booktabs}                  
\usepackage{lipsum}                    
\usepackage{mwe}                       

\usepackage{mathptmx}                  

\usepackage{enumitem}

\usepackage{amsmath}
\usepackage{amssymb}
\usepackage{bm}
\usepackage{newtxmath}

\usepackage{float}

\usepackage{bbm}

\usepackage{mathtools}

\usepackage{multirow}

\usepackage{siunitx}

\usepackage{url}

\newcommand{\sysname}{\textit{CausalPrism}\xspace}

\usepackage{xspace,xpunctuate}

\newcommand{\ie}{\textit{i.e.},\xspace}
\newcommand{\etal}{\xspace\textit{et al.}\xspace}
\newcommand{\eg}{\textit{e.g.},\xspace}
\newcommand{\etc}{\textit{etc.}\xspace}

\begin{document}



\maketitle

\section{Introduction}

\begin{figure}[tbp]
    \centering 
    \includegraphics[width=\columnwidth]{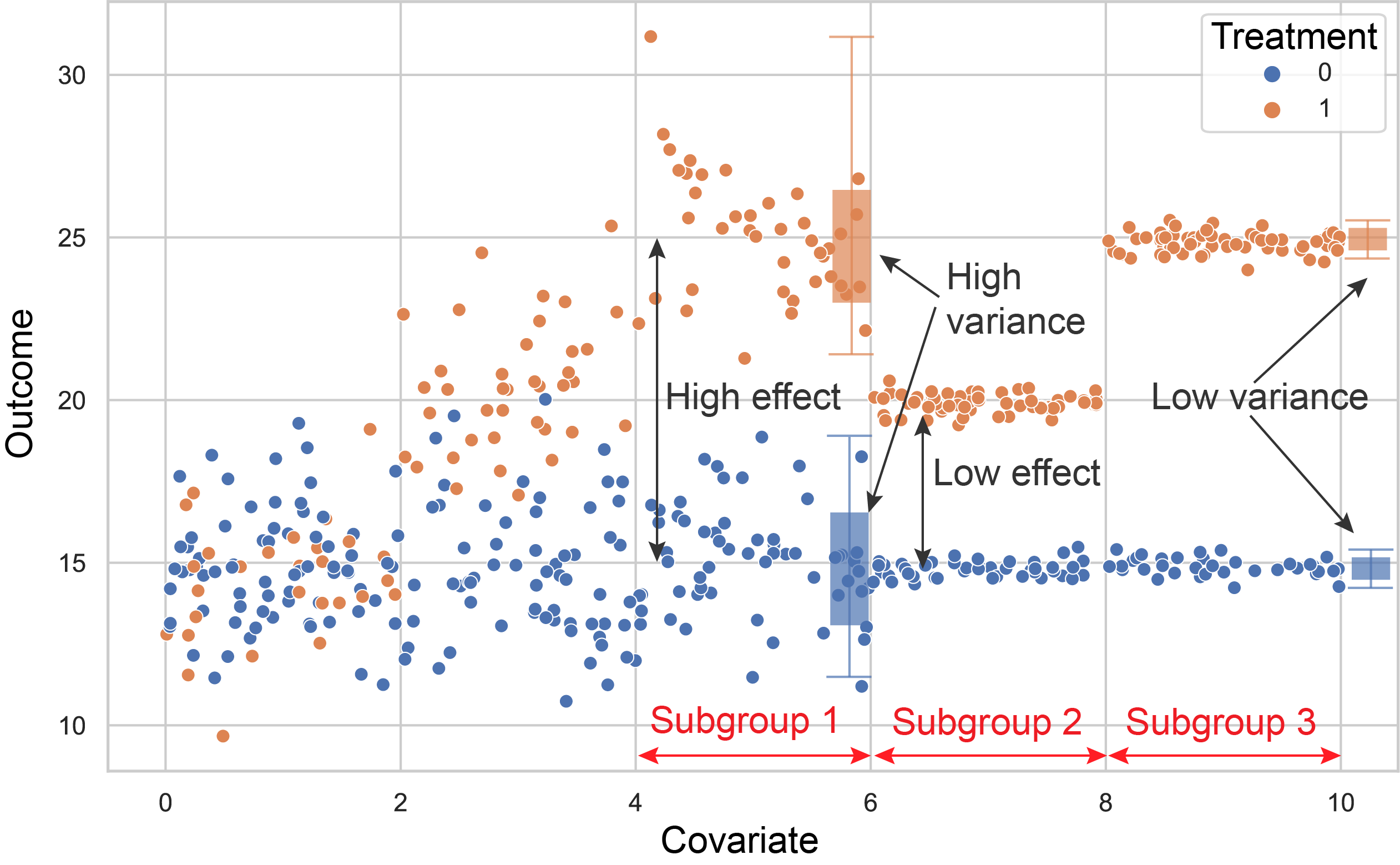}
    \caption{An illustrative toy example. There is only one covariate, and the change in the outcome between the treatment and control group can be informally thought of as the treatment effect. Subgroup 3 has a high effect and low variance, which is better than Subgroup 1 and 2.}
    \label{fig:causal-1}
\end{figure}

Causal inference is a data analysis process aiming at conclusions about whether and to what extent treatments affect outcomes~\cite{peters2017elements}. Data heterogeneity must be considered when estimating treatment effects, as the effect of the same treatment may vary across subgroups. As shown in \cref{fig:causal-1}, subgroups within the population responded differently to the treatment. The treatment exerts larger effects on Subgroup 1 and Subgroup 3 than Subgroup 2. Nevertheless, the high variance of Subgroup 1 indicates that individual differences (uncertainty of the outcome) exist. Discovering those subgroups with strong treatment effect and low outcome variance (hereinafter referred to as significant treatment effect) compared to the overall population is widely used in domains such as healthcare~\cite{rothman2005causation}, marketing~\cite{varian2016causal}, and public administration~\cite{gangl2010causal}. For example, marketers want to find customer groups where advertising more effectively drives purchases. Since Randomized Controlled Trials (RCTs), known as the gold standard for causal inference~\cite{Hariton2018rctgold}, are not always feasible due to cost or ethical concerns, there is a strong need to uncover those causal subgroups from observational data effectively.

In practical applications, analyzing causal heterogeneity faces two challenges. First, it is nontrivial and challenging to identify important subgroups from a large number of subgroup candidates. 
Subgroups can be described by different combinations of variables, which could lead to a combinatorial explosion of candidates. Selecting optimal subgroups requires trade-offs among various targets in the objective space, such as effect strength, outcome variance, and subgroup coverage, further complicating the subgroup discovery process. Coordinating the above factors has to follow users' analysis requirements. Nevertheless, the identification model can hardly communicate with humans due to the lack of interpretability, which causes the second challenge. 
The subgroups obtained by the black-box model may hardly be interpreted by or fail to support analysis tasks. Users still need to tediously analyze and compare multiple subgroups to determine which one they prefer. In addition, without a treatment effect explanation, numerical causal conclusions alone are difficult to convince users, especially in high-stakes safety and life-critical fields.

Existing work is insufficient to address these challenges. Automated heterogeneous treatment effect (HTE) estimation methods, such as causal trees~\cite{athey2016recursive} and causal forests~\cite{wager2018estimation}, mainly construct hierarchical structures for individuals in datasets and identify leaf nodes in the hierarchies as subgroups. However, not all leaf nodes can reflect a significant treatment effect, leading to less useful results. 
Therefore, users still need to go through tedious review and analysis to find subgroups that meet their requirements. Subgroup discovery methods~\cite{Atzmller2006SDMapA, Gamberger2002ExpertGuidedSD, Leeuwen2012DiverseSS} can directly optimize correlation objectives but lack attention to the more complex causal effects that require statistical inference and confounding bias correction.
Some researchers use visualization to assist causal analysis, but they either focus on the representation of causal graph structure~\cite{DBLP:journals/tvcg/WangM16, DBLP:journals/tvcg/XieDW21, DBLP:conf/ieeevast/WangM17, DBLP:journals/tvcg/JinGCWGC21}, manual selection of variables to divide subgroups~\cite{10.1145/3544548.3581236}, or clustering to obtain subgroups whose meaning is difficult to describe~\cite{9623285}. Therefore, how to support the cooperation between human intelligence and computing power in the analysis of HTE is still underexplored.

In this work, we propose a visual analytics approach for subgroup-based causal heterogeneity exploration that supports users in effectively identifying optimal subgroups from observational data, comparing and ranking different subgroups, and verifying treatment effects. First, we propose a causal subgroup discovery model based on constrained multi-objective optimization (MOO). Subgroups are described by interpretable rules, where the rule antecedents are conjunctions containing covariates and corresponding values, and the consequents correspond to subgroup evaluation metrics, such as effects and variances. Coverage and antecedent length are used as constraints to ensure interpretability. Due to multiple objectives and constraints, the optimal Pareto front of the subgroup is learned by a heuristic searching algorithm for user analysis. Second, we design and develop an interactive prototype system, \sysname, which incorporates intuitive visualizations of subgroups, evaluation metrics, and explanations of treatment effects, thereby facilitating users' understanding, comparison, and verification of causal subgroups. Quantitative experiments, case studies, and expert interviews demonstrate the effectiveness and usability of the proposed model and system. In summary, our contributions are as follows:

\begin{itemize}[noitemsep,topsep=0pt]
    \item We propose a subgroup discovery model based on constrained multi-objective optimization, which can mine rule-explained subgroups with significant treatment effects from a large amount of high-dimensional observational data that outperforms state-of-the-art methods.
    
    \item We designed and implemented an interactive visual analytics prototype system, \sysname, which includes table-based subgroup visualization, multi-attribute ranking, and matching unit-based explanation of treatment effects. The system supports users in understanding, comparing, and validating causal subgroups. Its utility has been proved through case studies, and positive feedback has been received during expert interviews.

\end{itemize}
\section{Related Work}

\subsection{Heterogeneous Treatment Effect Estimation}

Treatment effects vary across the whole population. Conditional average treatment effect (CATE), individual average treatment effect (ITE) and causal rules comprise current HTE research. Reviews~\cite{10.1145/3444944, 10.1145/3397269} offer in-depth analyses of causal inference.

\textit{CATE} evaluates treatment effects on specific subgroups of the population, given similar covariates like demographics. To optimize the heterogeneity of treatment effects, tree-based methods~\cite{athey2016recursive, wager2018estimation, athey2019estimating} are commonly employed to partition the covariate space into subspaces. For instance, Causal Tree~\cite{athey2016recursive} constructs the tree and estimates treatment effects in each subspace using separate data, avoiding overfitting by cross-validation. Wager\etal~\cite{wager2018estimation} suggested Causal Forest, combining causal tree ensemble results for more robust and smooth estimation. Root-to-leaf node routes naturally define subgroups of heterogeneous CATEs, making the tree model interpretable. However, tree-based methods may have limited performance due to greedy tree building process and does not necessarily return the ``optimal'' structure.

\textit{ITE} compares outcomes with and without treatment. Since only one outcome is visible, another must be estimated. Existing techniques are single- or multi-model-based depending on whether treatment and control groups are estimated independently. The former uses regression to fit treatment effects. For example, Hill\etal~\cite{hill2011bayesian} employs Bayesian additive regression trees to fit the outcome surface. The latter fits the treated and control groups separately, achieving better performance for significant differences between groups' outcomes. The base model uses off-the-shelf estimators like linear regression~\cite{cai2011analysis} or neural networks~\cite{johansson2016learning}. With well-tuned parameters, these models can accurately estimate effects but are uninterpretable.

Several researchers have tried to find \textit{causal rules} in data. For example, CRE~\cite{Lee2020CausalRE} is a two-stage method that first produces rules using methods such as random forest or Gradient Boosting Machines, then picks robust ones using stability selection regularization. After mining association rules from data, Li\etal~\cite{Li2015FromOS} conducted a cohort study to test whether the association rules were causal. 

Many methods have presented estimators that can accurately estimate HTE despite confounding biases. We leverage existing estimators but focus on finding subgroups with significant treatment effects among many candidates. We present a constrained MOO-based causal subgroup discovery model that outperforms tree-based and black-box approaches in significance and interpretability.

\subsection{Visual Causality Analysis}

Automated causal detection algorithms are built upon assumptions and complex causal mechanisms that are hardly fulfilled in real life, causing accuracy and interpretability issues. Visualizations have been used to explore and verify causality interactively, which can be divided into homogeneous and heterogeneous causality investigations.

\textit{Homogeneous} causality assumes that data's causal mechanism is static and stable.
Visualization helps users grasp complex causal relationships and make decisions. 
Graph-based visualizations~\cite{DBLP:journals/tvcg/WangM16, DBLP:journals/tvcg/XieDW21, 9507307, DBLP:journals/cgf/BaeHR17} have been widely used to demonstrate causality in multi-attribute datasets, emphasizing the use of statistics to locate and manipulate improper relationships for what-if analysis. 
They employ advanced layout designs to highlight attribute distributions and enhance graph readability.
In addition, design factors like node size~\cite{DBLP:journals/tvcg/KadabaIL07}, edge shape~\cite{DBLP:conf/grapp/BaeVRHF17}, and crowd beliefs~\cite{10.1145/3544548.3581021} affect users' understanding of causal relationships. 
Bar charts and scatter plots are also used to infer causality~\cite{DBLP:journals/tvcg/XiongSHF20, DBLP:journals/cgf/YenPF19, DBLP:journals/tvcg/KaleWH22}. 
However, these approaches lack generalizability due to population variances such as demographics and environmental factors.

\textit{Heterogeneous} causality examines causal relationships or effects that vary over time or data subgroups. Most work examines causal structural heterogeneity. In Causal Structure Investigator~\cite{DBLP:conf/ieeevast/WangM17}, users can acquire data subdivisions through manual filtering and k-means clustering. Then, these subdivisions are mapped to causal graphs for detailed analysis of causal paths. Jin\etal~\cite{DBLP:journals/tvcg/JinGCWGC21} focus on subsets in event sequences. Overlapping adjacency matrices with inner and outer sections lets users easily identify the differences in causal relationships between subsets. 
Deng\etal~\cite{DBLP:journals/tvcg/DengWXBZXCW22} created causal graph bands with compass glyphs for spatio-temporal sequences to show dynamic causal relationships in period-based time windows. This helps users understand influence transmission and identify spurious causalities. DOMINO~\cite{wang2022domino} applies time delays and event constraints to temporal causality analysis, facilitating hypotheses formulation and validation. 


Other research examines causal inference heterogeneity. The Absolute Standardized Mean Difference (ASMD) plot is used to assess covariate balance in groups after weighting and propensity score matching~\cite{Shimoni2019AnET, cobalt}. Guo\etal~\cite{9623285} created VAINE to enable users to find statistical phenomena like Simpson's paradox by manually selecting clusters in covariate projections and observing their impact. Causalvis~\cite{10.1145/3544548.3581236}, a later proposal, enables the visualization of a whole causal analysis workflow. The raincloud and beeswarm plots in the Treatment Effect Explorer module let users manually pick subgroups faceted by covariates and analyze ITE distribution to examine heterogeneity.

However, the present HTE visualization work involves time-consuming manual participation to locate subgroups, and the subgroups obtained through clustering lack explicit interpretable descriptions. Therefore, we propose \sysname to automatically obtain rule-described subgroups with significant treatment effects through optimization and design visualizations for subgroup exploration, comparison, and treatment effect validation.

\subsection{Subgroup Discovery and Visualization}


Subgroup discovery (SD) is a descriptive data mining method that finds data subgroups with intriguing patterns on certain goals, as summarized in comprehensive surveys~\cite{Herrera2011AnOO, AtzmuellerSubgroupD}.
Data subgroups can be represented using description languages like attribute-value pairs and logical forms(\eg conjunctions, inequalities, and fuzzy logic). Subgroup interestingness can be measured using binary, nominal, or numerical targets. Post-processing methods have been applied to select diverse and less redundant subgroups. Search methodologies like exhaustive and heuristic search have been used due to the large number of candidate subgroups.

Using the exhaustive techniques~\cite{Wrobel1997AnAF, Atzmller2006SDMapA, Grosskreutz2008TightOE, Grosskreutz2009OnSD}, all possible subgroups are searched. Since viable subgroups are exponentially large, a naive exhaustive search is time-consuming. Minimum support, optimistic estimate pruning, and generalization-aware pruning can reduce the hypothesis space. SD-Map~\cite{Atzmller2006SDMapA} is an exhaustive SD approach that uses depth-first search to produce candidates, extending the Frequent Pattern (FP) Growth-based association rule mining method. 
The SD-Map*~\cite{Atzmller2009FastSD} is extended with binary, categorical, and continuous target variables.

Further studies~\cite{Gamberger2002ExpertGuidedSD, Lavra2004SubgroupDW, Leeuwen2012DiverseSS, Jess2007EvolutionaryFR, zitzler2001spea2} employed efficient heuristic methods. For example, DSSD~\cite{Leeuwen2012DiverseSS} uses beam search, which starts with an initial solution and subsequently spreads to several candidates. Top performers are kept for the next iteration until a stopping condition is reached. SDIGA~\cite{Jess2007EvolutionaryFR} is an evolutionary fuzzy rule induction method that facilitates the discovery of general rules by allowing variables to take multiple values. Subgroups can be evaluated in terms of confidence, support, and unusualness.

Visualization techniques have also been proposed in order to support subgroup-level analysis tasks, such as subgroup multi-feature visualization~\cite{6634146, doi:10.1177/1473871619878085}, model diagnosis on data subsets~\cite{10292663, 9229232, 8464305, 9973226, 9906903}, and high-dimensional data subspace exploration~\cite{10443294, https://doi.org/10.1111/cgf.14290, 8017645}. For example, Taggle~\cite{doi:10.1177/1473871619878085} employs a tabular visualization design that allows for hierarchical grouping and sorting of massive amounts of data. The icicle plot~\cite{8464305} and the map-based metaphor~\cite{10443294} provide help for comparisons between subgroups.

However, most SD methods only focuse on correlations, involving just covariates and outcomes. It is unsuitable for SD in causal scenarios (treatment, covariates, and outcomes must be considered). To this end, we formulate causal SD as a constrained MOO problem that can be efficiently solved using heuristic search. A range of subgroup visualization techniques, such as multi-attribute ranking~\cite{6634146}, are incorporated into the \sysname system to help users explore and compare subgroups.
\section{Background}

\subsection{Preliminaries}

We introduce the basis of causal inference under the potential outcome (PO) framework~\cite{10.1093/oxfordhb/9780199286546.003.0011} and give examples based on medical scenarios.

A \textbf{unit} is an individual or object under study. A medical study unit may be a patient. The subscript ${}_{i}$ denotes the $i$-th unit.

A \textbf{treatment} is an intervention or exposure that subjects to a unit. A new medicine or therapy could be used as a treatment in a medical study. Let a binary $T$ indicate whether a unit has received a treatment. Units satisfying $T=1$ belong to the treatment group, while those $T=0$ belong to the control group. 

\textbf{Outcomes} are what would happen to units under different treatments. Each unit has two potential outcomes: factual outcome and counterfactual outcome. For instance, patient survival time is an outcome in a medical study. The potential outcome with treatment is $Y(T=1)$, also abbreviated as $Y(1)$, and without treatment, it is $Y(0)$.

\textbf{Covariates} are background variables that affect treatment assignment and outcome. For example, patient demographic information such as age may influence medication use (treatment assignment) and blood pressure (outcome). Observational studies often control for covariates to mitigate confounding and provide more unbiased effects estimates. Covariates are represented as a vector $\mathbf{X}_i=(x_{i,1},\cdots,x_{i,d})$, where $d$ is the number of covariates.

\textbf{Observational data} refers to data collected without the researcher manipulating the environment or the subjects being studied. It differs from RCTs, which randomly assign treatment to each unit. The observational data containing $n$ units is denoted by $\mathcal{D} = \{(T_i,\mathbf{X}_i,Y_i)\}_{i=1}^n$.

\textbf{Treatment effect} refers to the impact of a treatment on an outcome. It can be obtained by quantitatively comparing the potential outcomes in the treatment and control conditions at different levels, such as populations, subgroups, and units. For unit $i$, its individual treatment effect (ITE) is defined as:

\begin{equation}
\tau_i=Y_i(1)-Y_i(0).
\end{equation}

Unfortunately, for any unit, only one of the two potential outcomes can be observed, so ITE is not identifiable. One way to address this lack of counterfactual outcomes is to estimate the average treatment effect (ATE) on the population, defined as follows: 

\begin{equation}
\tau=\mathbb{E}[Y(1)-Y(0)].
\end{equation}

ATE may fail to accurately reflect treatment effects due to the heterogeneity of units. This is overcome by conditional average treatment effect (CATE) on subgroups, which is defined as follows:

\begin{equation}
\tau(\mathbf{x})=\mathbb{E}[Y(1)-Y(0) \mid \mathbf{X}=\mathbf{x}].
\end{equation}

\textbf{Propensity score} is a balancing score $e(\mathbf{x})=P(T=1\mid \mathbf{X}=\mathbf{x})$, defined as the conditional probability of getting a treatment given the covariates. In observational data, a biased treatment effect would be obtained by directly using the difference between the average outcome of the treatment and control groups because the treatment assignment is correlated with covariates. Rubin \etal~\cite{Blumberg2016CausalIF} proved that $\{Y(0), Y(1)\} \perp T \mid e(\mathbf{X})$ under the assumption of unconfoundedness. For binary treatment, the Logistic regression model is commonly used to estimate propensity scores~\cite{Caliendo2005SomePG}.

\textbf{Outcome Variance} refers to the outcome variability among the units in the treatment and control group. A lower variance means that the treatment leads to a more consistent outcome among units. Let $\sigma^2$ be the variance.

\textbf{Causal Subgroup} refers to specific subgroups within the population that exhibit significant treatment effects. For example, the preventive effect of influenza vaccine is more significant in the elderly and immunocompromised people. $\mathcal{S}$ is used to represent subgroups.

\subsection{Design Requirements}\label{sec:req}

We distilled the design requirements from interviews with three experts (E1-3) and a literature review. Data analysts E1 and E2 have three years of work experience in a technology firm. Their daily tasks include evaluating KPI anomalies and guiding advertising placement using causal analysis on observational data. E3, a university associate professor, has written multiple causal inference studies. They noted that causal inference is plagued by data heterogeneity, and massive amounts of observational data lack appropriate exploration tools. Causality interpretability is also crucial since users cannot make decisions if they don't trust the result. Finally, the requirements are listed.


\begin{enumerate}[label=\textbf{R{\arabic*}}, nolistsep]
    \item
    \textbf{Descriptive subgroup identification}. Observational data usually contains a large potential exploration space with many variables. Although traditional data clustering methods can be used to discover clusters, they do not give a corresponding interpretation. Experts mentioned that ``Although high-value groups can be manually segmented based on domain knowledge, it often requires multiple attempts of different filtering conditions.'' Users can take advantage of automatically identified subgroups to further discover interesting causal patterns.

    \item
    \textbf{Subgroup understanding and valuation}. Subgroups involve rich information such as variables used in subgroup descriptions, value distribution of covariates, treatment effect, and variance scores. Users should be able to browse such information to understand the characteristics of a subgroup. It is necessary for the system to provide a clear and intuitive visualization for subgroups.

    \item
    \textbf{Subgroup adjustment and hypothesis}. The subgroups automatically discovered by the model may not satisfy users. Experts say that for advertising scenarios, they aim to boost user spending and meet total profit goals. Therefore, when necessary, they will relax the filtering conditions or merge small groups to enlarge the target subgroup. Our approach needs to support users to adjust subgroups. The basis of adjustment could be the understanding of the target subgroup, domain knowledge from users, or the analysis results of what-if tests on new subgroups.

    \item
    \textbf{Subgroup comparison and ranking}. Users have diverse preferences for causal subgroups. For example, when determining the target audience for advertising, conservative users are willing to choose subgroups that are generally effective and have smaller variances, while risk-takers try subgroups that have stronger effects but may also have greater outcome fluctuations. Therefore, we need to allow users to compare subgroups from multiple perspectives and rank them based on their preferences to select satisfactory subgroups.

    \item
    \textbf{Treatment effect validation}. Users need to understand why certain treatment effect is estimated and be provided with data evidence to explain their rationality. In addition, the estimated treatment effect may be biased by the size and distribution of the data units. Seeking reliable conclusions, visualizations are needed to help users rule out suspicious causal effects.

\end{enumerate}

\section{Workflow Overview}\label{sec:workflow}

To address the requirements mentioned in \cref{sec:req}, we developed \sysname, a visual analytics system for analyzing causal heterogeneity from a subgroup perspective, allowing users to identify, explore, rank, and validate causal subgroups. \Cref{fig:workflow} illustrates the workflow of \sysname. The input is observational data, including covariates (\eg age, height, weight), a treatment (\eg drug), and an outcome (\eg survival time). Given the input data, the system works as follows:

\begin{figure*}[ht!]
    \centering 
    \includegraphics[width=2\columnwidth]{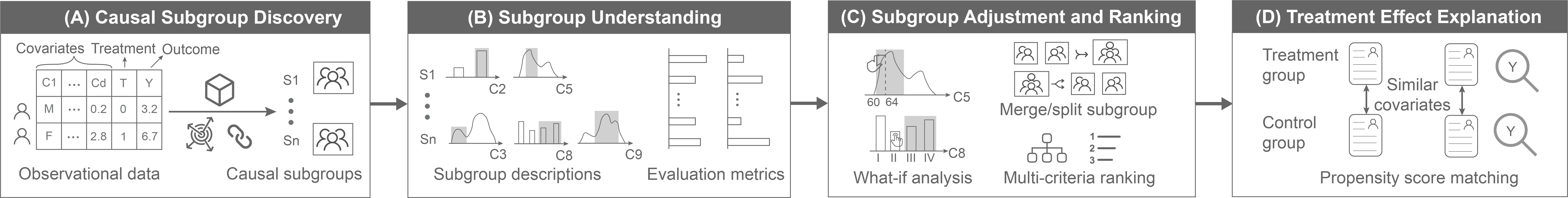}
    \caption{A four-step workflow for subgroup-based causal heterogeneity exploration. (A) The model automatically mines subgroups with significant treatment effects from observational data. (B) Subgroups can be explored through tabular and multi-attribute visualizations. (C) Users can interactively analyze new subgroup hypotheses and achieve multi-criteria decision-making based on their preferences. (D) Effect interpretation based on matched units simulates the user's familiar A/B testing, aiding in result validation.}
    \label{fig:workflow}
\end{figure*}

(A) First, causal subgroups are automatically identified by the model from the observational data in which the treatment has a significant effect on the outcome. Since there are multiple objectives, including treatment effects and outcome variance, the discovered subgroups are usually an optimal solution set. Constraints such as subgroup coverage can also be imposed on the subgroup discovery process. (\textbf{R1}, \cref{sec:subgroup-model})

(B) Second, to understand the subgroups, their interpretable descriptions (\eg ``age>60 AND sex=female'' can be considered as a description of the older female group) and evaluation metrics (\eg treatment effects, outcome variance) are presented. Users can get an overview of the discovered subgroups and view the details of a subgroup on demand. (\textbf{R2}, \cref{sec:subgroup-view})

(C) Third, custom subgroups are allowed, where users can hypothesize subgroups based on domain knowledge for what-if analysis. By weighing evaluation metrics such as effect strength and subgroup size, users can make multi-criteria decisions and select a preferred subgroup. (\textbf{R3}, \textbf{R4}, \cref{sec:subgroup-view}, \cref{sec:projection-view})

(D) Finally, after the subgroups of interest have been selected, explanations of the treatment effect within the subgroups can be examined. Analytical evidence, such as the uncertainty of individual treatment effects, is provided, thus helping users exclude the unreliable subgroup and enhance trust in the results. (\textbf{R5}, \cref{sec:explanation-model}, \cref{sec:validation-view} )
\section{Models} \label{sec:model}

In this section, we introduce the models used in our system.

\subsection{Causal Subgroup Discovery} \label{sec:subgroup-model}

We first frame causal subgroup mining as a constrained multi-objective optimization problem and solve the optimal subgroup set through an efficient heuristic algorithm. (\textbf{R1})

\subsubsection{Problem formulation}

Without loss of generality, we consider observational data whose covariates are binary, \ie $\mathbf{X}_i=(x_{i,1},\cdots,x_{i,d})\in\{0,1\}^d$. Categorical variables can be binarized by one-hot encoding (such as color=red, color=yellow, color=blue). As for numerical variables, we can convert the value intervals to binary by bucketing strategy (such as age$\le$10, age>10, $\cdots$, age$\le$100, age>100). Formally, our goal is to learn causal subgroups $\mathcal{S}$ from the given observational data $\mathcal{D}$. We use interpretable rules (simple logical structures of the form ``IF $P$ THEN $Q$'') to describe the subgroup $\mathcal{S}: \bm{\alpha} \Rightarrow [\tau, \sigma^2(0), \sigma^2(1)]$, which contain the antecedent $\bm{\alpha}$ and the consequent $[\tau, \sigma^2(0), \sigma^2(1)]$.

A \textbf{antecedent} $\bm{\alpha}$ is the condition of the subgroup, expressed as the conjunctive normal form (CNF) of a series of atoms $\bigwedge_{j \in \Gamma} x_{j}$, \eg ``age > 25 AND job == teacher''. $\Gamma$ is the covariate indices used in the antecedent, which is a subset of the indices of all binary covariates, \ie $\Gamma \in 2^{[d]}$, where $[d] = \{1, \cdots, d\}$ and $2^{[d]}$ means the power set of $[d]$. The atom $x_j$ is the smallest interpretable element.
The mapping from a $\mathcal{S}$ to a CNF is given by $\bm{\alpha}_{\mathcal{S}}(\mathbf{X}_i) = \bigwedge_{j \in \Gamma_{\mathcal{S}}} x_{i,j}$. For brevity, we call it as $\bigwedge_{j \in \mathcal{S}} x_{i,j}$. When $\bm{\alpha}_{\mathcal{S}}(\mathbf{X}_i)$ is true, the $i$-th unit is \textbf{covered} by the subgroup $\mathcal{S}$. We define the length of the antecedent $|\bm{\alpha}|$ as the number of different covariates. For example, the length of the antecedent ``10 < age <= 25 AND sex = female'' is 2. $|\bm{\alpha}|$ reflects the readability of the antecedent; shorter antecedents are easier for users to interpret.

The \textbf{consequent} is the evaluation results of the subgroup, consisting of the treatment effect $\tau$, the control group variance $\sigma^2(0)$, and the treatment group variance $\sigma^2(1)$. 

To estimate the CATE $\tau$ for units covered by the subgroup, we employ Inverse Probability Weighting (IPW)~\cite{Hirano2000EfficientEO}, which assigns appropriate weights $w_i=\frac{T_i}{e_i} + \frac{1-T_i}{1-e_i}$ to each unit to balance the distribution of covariates in the treatment and control groups, thereby simulating RCTs. The normalized weighted average of the factual outcomes for the treatment and control groups can estimate treatment effects~\cite{10.1162/003465304323023651}:

\begin{equation}
    \tau_{\mathcal{S}} = \dfrac{\sum_{i \in \mathcal{D}_{\mathcal{S}}^+} w_iY_i}{\sum_{i \in \mathcal{D}_{\mathcal{S}}^+} w_i} - \dfrac{\sum_{i \in \mathcal{D}_{\mathcal{S}}^-}w_iY_i}{\sum_{i \in \mathcal{D}_{\mathcal{S}}^-}w_i},
\label{eq:effect}
\end{equation}

where $\mathcal{D}_{\mathcal{S}}$ denotes the covered data, $\mathcal{D}^+$ denotes the data that received the treatment ($T=1$), and $\mathcal{D}^-$ denotes the data that did not receive the treatment ($T=0$), $\mathcal{D}_{\mathcal{S}}^+ = \{ i | i \in \mathcal{D}^+ \wedge \bm{\alpha}_{\mathcal{S}}(\mathbf{X}_i) = 1 \}$ denotes the units in the treatment group that are covered by the subgroup $\mathcal{S}$, $\mathcal{D}_{\mathcal{S}}^- = \{ i | i \in \mathcal{D}^- \wedge \bm{\alpha}_{\mathcal{S}}(\mathbf{X} _i) = 1 \}$ denotes the units in the control group that are covered by the subgroup $\mathcal{S}$. 

Since obtaining treatment effects is a statistical estimation problem, it is important to consider the uncertainty of the treatment effect, which can be measured by the outcome variance, defined as:

\begin{equation}
\begin{split}
    \sigma_{\mathcal{S}}^2(0) & =\dfrac{\sum_{i \in \mathcal{D}_{\mathcal{S}}^-} w_i(Y_i - \overline Y_w)^2}{\sum_{i \in \mathcal{D}_{\mathcal{S}}^-} w_i} \\
    \sigma_{\mathcal{S}}^2(1) & =\dfrac{\sum_{i \in \mathcal{D}_{\mathcal{S}}^+} w_i(Y_i - \overline Y_w)^2}{\sum_{i \in \mathcal{D}_{\mathcal{S}}^+} w_i},
\end{split}
\label{eq:variance}
\end{equation}

where $\overline Y_w$ is the weighted outcome mean.

Therefore, we formalize learning causal subgroups from observational data as a \textbf{constrained multi-objective optimization} problem:

\begin{equation}
\begin{aligned}
\label{eq:opt}
\max_{\mathcal{S}} \; & \tau_{\mathcal{S}} \\
\min_{\mathcal{S}} \; & \sigma_{\mathcal{S}}^2(0), \sigma_{\mathcal{S}}^2(1) \\
\text {s.t.} \; & |\mathcal{D}_{\mathcal{S}}| \geq C \\
& |\bm{\alpha}_{\mathcal{S}}| \leq L.
\end{aligned}
\end{equation}

To ensure interpretability and meet user personalized needs, two constraints are added. $|\mathcal{D}_{\mathcal{S}}| \leq C$ limits the unit covered by the subgroup to at least $C$, and $|\bm{\alpha}_{\mathcal{S}}| \leq L$ limits the length of the antecedent by $L$.

\subsubsection{Solving the problem}

\begin{figure}[tbp]
    \centering 
    \includegraphics[width=\columnwidth]{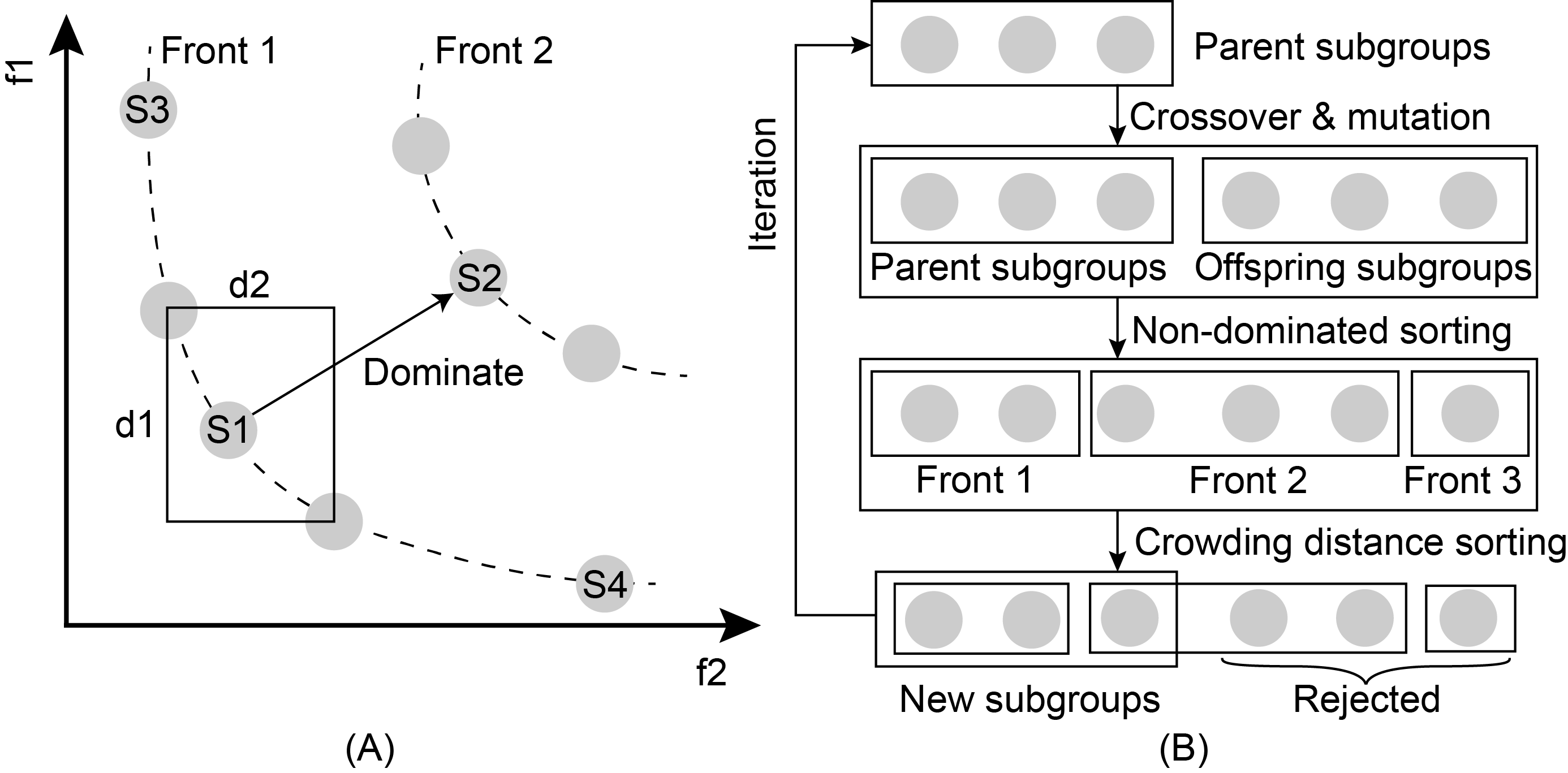}
    \caption{Illustration of the proposed model to discover optimal causal subgroups. (A) Schematic diagram of the Pareto front, where circles represent feasible subgroups for which small objective values are preferred over large values. The subgroup in Front 1 is not dominated by other subgroups, and the subgroup in Front 2 is dominated only by those in Front 1. (B) Illustration of the iterative heuristic algorithm for solving the multi-objective optimization problem. In each iteration, offspring subgroups are generated from the parent subgroup and survived by first comparing the front level and then the crowding distance.}
    \label{fig:causal-SD}
\end{figure}

Solving the optimization problem of \cref{eq:opt} is not easy because the decision variable $\mathcal{S}$ is a rule-described subgroup rather than a single continuous variable, and the problem contains multiple objectives and constraints that are difficult to differentiate. Analytic solution or gradient descent methods are thus not applicable to this problem. For multi-objective problems, it is often impossible to obtain a single ideal optimal solution because optimizing one objective is likely to be at the expense of another objective. Therefore, the solution most often consists of a series of non-dominated subgroups, more formally called the Pareto front. As shown in \cref{fig:causal-SD}-(A), assuming two minimizing objectives $f_1$ and $f_2$ (if there are both maximizing and minimizing objectives, we can convert them to minimizing by adding a negative sign to maximizing), the circle represent a feasible solution (\ie a subgroup in our problem). Front 1 denotes the set of non-dominated solutions because no other solution is better than them. In mathematical terms, one solution $s_1$ (Pareto) dominates another solution $s_2$, if

\begin{equation}
\begin{aligned}
& \forall i \in\{1, \ldots, m\}, f_i\left(s_1\right) \leq f_i\left(s_2\right), \text { and } \\
& \exists i \in\{1, \ldots, m\}, f_i\left(s_1\right)<f_i\left(s_2\right) ,
\end{aligned}
\end{equation}

where $m$ is the number of objectives.

In order to efficiently find the subgroups belonging to the Pareto front, we employ a heuristic genetic search algorithm~\cite{996017}. As shown in \cref{fig:causal-SD}-(B), the algorithm mainly consists of 4 steps.

1. \textbf{Subgroups initialization}. We use binary random sampling to generate binary vectors, where 0 represents not selecting the corresponding covariates, and 1 indicates that it is selected, resulting in different antecedents for describing the subgroups. The default initial number of subgroups is 100.

2. \textbf{Offspring generation}. Existing subgroups are used to generate new subgroups by crossover and randomly flipping binary vectors. For example, a new subgroup is described by splicing the first half of the antecedent in Subgroup $S_1$ onto the second half of the antecedent in Subgroup $S_2$. This step helps to expand the search space as much as possible without falling into a local optimum. Together, the existing and new subgroups form a candidate population.

3. \textbf{Non-dominated sorting}. The dominance relationship between all subgroups is obtained through a pairwise check. The subgroups that are not dominated by any other subgroups form Pareto front 1; the subgroups dominated only by the front 1 are the front 2, and so on. Therefore, all subgroups are divided into several ordered levels.

4. \textbf{Crowding distance sorting}. The crowding distance is the Manhatten distance in the objective space, which measures the distribution density of subgroups. For instance, in \cref{fig:causal-SD}-(A), the crowding distance of subgroup $S1$ is $\frac{d_1}{f_1^{\max} - f_1^{\min}} + \frac{d_2}{f_2^{\max} - f_2^{\min}}$. The boundary subgroups in front (such as $S3$ and $S4$) will be given an infinite crowding distance. The subgroup with a larger crowding distance is given priority, thereby ensuring the diversity of the solution space. After sorting with the front as the first priority and crowding distance as the second priority, the top-ranked subgroups are retained, while other subgroups are discarded.

Steps 1-4 are performed iteratively until a predetermined number of iterations is reached or the target values of the subgroups no longer improve. During this process, solutions that do not satisfy the constraints or are duplicated are eliminated. The subgroups belonging to front 1 obtained at the end of the iterative process are used as outputs.

\subsection{Treatment Effect Explanation} \label{sec:explanation-model}

Unlike prediction or classification problems, the ground truth of treatment effects is often unknown because we cannot simultaneously observe two contrasting outcomes. Existing causal inference methods usually estimate treatment effects through confounding balancing and statistical inference, which are naturally uncertain. Therefore, providing reasonable explanations for treatment effects is crucial to help users interpret the results and enhance trust. (\textbf{R5})

\begin{figure}[tbp]
    \centering 
    \includegraphics[width=\columnwidth]{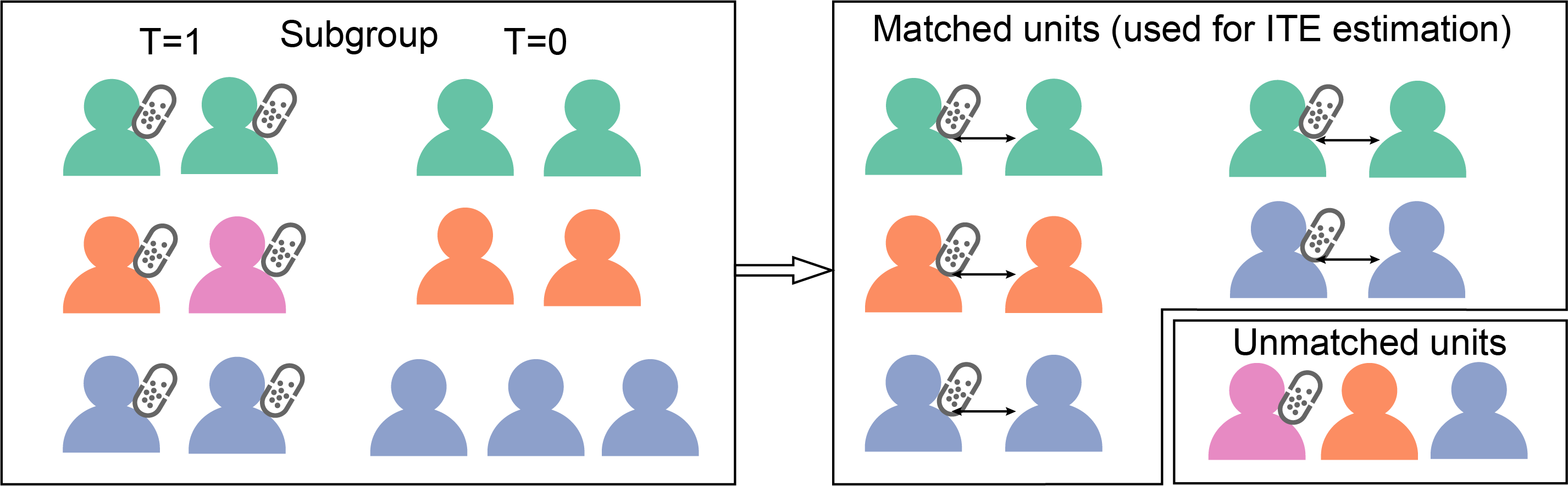}
    \caption{Illustration of propensity score matching. Colors represent different covariates, and drug icons indicate treatment. Matching reduces confounding bias by finding comparable treatment and control units.}
    \label{fig:causal-Exp}
\end{figure}

We utilize propensity score matching to generate explanations of treatment effects, which mimic the A/B testing familiar to users. As shown in \cref{fig:causal-Exp}, for subgroups automatically discovered or manually added by users, the units covered may have different covariates. For example, the elderly may have different demographic characteristics such as height and weight. To balance these confounders, we identify pairs of units from treatment and control groups whose covariates are similar or even identical, \ie $\text{Distance}(\mathbf{X}_i, \mathbf{X}_j) \le \epsilon$. The naive matching method is exactly matching, which requires that the matched unit covariates are identical; however, in high-dimensional settings, there are rarely exact matches. Therefore, we match units by the distance between propensity score, \ie $\text{Distance}(\mathbf{X}_i, \mathbf{X}_j)=|\hat{e}(\mathbf{X}_i) - \hat{e}(\mathbf{X}_j)|$, which transforms the problem of high-dimensional space matching into scalar propensity score matching and controls for confounding bias. In the implementation, we set the threshold $\epsilon=0.1$ and greedily find the nearest matching unit in the control group for the units in the treatment group, guaranteeing that each unit matches at most once. Based on the difference in the outcomes (individual treatment effects) for the matched units, users can learn about the concentration and distribution of effects for that subgroup and thus judge the reliability of the estimated effects.
\section{Interface Design}

We developed a prototype system, \sysname, to help users implement the workflow proposed in \cref{sec:workflow}. This section presents an overview of the system and details of visual design and interaction.

\subsection{System Overview}

As depicted in \cref{fig:teaser}, \sysname offers three views: Causal Subgroup, Covariate Projection, and Treatment Effect Validation. These views enable subgroup exploration, comparison, and explanation of treatment effects. We demonstrate an analysis flow that uses these views to analyze causal heterogeneity in observational data. A data analyst wants to see if the treatment affects the outcome differently across subgroups. She enters the data into the system and sets treatment, outcome, maximum antecedent length, and minimum coverage. The model (\cref{sec:subgroup-model}) automatically identifies subgroups with significant treatment effects (\textbf{R1}). The Causal Subgroup View (\cref{fig:teaser}-A) offers an overview of causal subgroups, including their description and evaluation metrics (\textbf{R2}). Based on domain knowledge, she adds a new subgroup to analyze together (\textbf{R3}). In the Covariate Projection View (\cref{fig:teaser}-B), she integrates small, close subgroups into a large one (\textbf{R3}). Since there are many possible subgroups, the multi-attribute ranking function is resorted to identify preferred ones (\textbf{R4}). To better understand the treatment effect in the subgroup, she consults the Treatment Effect Validation View (\cref{fig:teaser}-C). The propensity score histogram shows the proportion of matched units in treatment and control groups. She then interprets the effect strength and uncertainty using dot plots of matched pairs and detailed information (\cref{sec:explanation-model}, \textbf{R5}).

\subsection{Causal Subgroup View} \label{sec:subgroup-view}

The Causal Subgroup View contains three parts: (1) a table-based antecedent visualization to intuitively convey the meaning of the subgroup (\textbf{R2}); (2) a subgroup editing window for additions and modifications to improve the identified subgroups (\textbf{R3}); and (3) an evaluation metrics visualization that supports multi-attribute ranking to facilitate personalized comparison of subgroups (\textbf{R4}).

Users need antecedents and consequents to grasp causal subgroups. Antecedents include covariates and values in the form of CNF, and consequents include evaluation metrics. Thus, various one-to-many relationships (subgroup $\rightarrow$ covariates, covariate $\rightarrow$ values, subgroup $\rightarrow$ metrics) exist. Tabular, matrix, and multi-attribute ranking visualizations inspire us to intuitively visualize this information. Tabular/matrix forms like UpSet~\cite{6876017} and Taggle~\cite{doi:10.1177/1473871619878085} are useful for analyzing set relationships. Multi-attribute ranking visualizations like LineUp~\cite{6634146} and SRVis~\cite{8456575} aid in multi-criteria decision-making. Our bespoke table-based view(\cref{fig:teaser}-A1) shows subgroups as rows and covariates and evaluation metrics as columns. Expanding a covariate column displays its distribution. Bar charts display distinct value counts for category covariates. Smooth line charts show the count over numerical covariates' value domain. The cell displays subgroup covariate values as circles or rectangles, reflecting discrete values and continuous intervals. Clicking the ``Add Subgroup'' or ``Edit Subgroup'' button opens the ``Subgroup Edit Box''(\cref{fig:teaser}-A2), allowing users to alter covariate values and identify the subgroup with relevant domain semantics.


The right side of the table (\cref{fig:teaser}-A3) displays a multi-attribute depiction of subgroup evaluation metrics, with each column representing a measure. Horizontal bars represent metric values. Users can apply an inverted mapping to measures like variance, where more minor is better. Users can drag to combine metrics and add weights to construct stacked bar charts to convey preferences. Sorting and filtering lets users find relevant subgroups faster.

\subsection{Covariate Projection View} \label{sec:projection-view}


The Covariates Projection View (\cref{fig:teaser}-B) displays each unit's low-dimensional projection. Users have trouble intuitively comparing units because raw observations include many covariates. To downscale high-dimensional units to the 2D plane, we use non-metric multidimensional scaling~\cite{kruskal1964nonmetric}, where similar units are close. The scatter plot highlights units from the selected subgroup in the Causal Subgroup View, showing subgroup size and connectivity. To adjust subgroup size, users can click ``Merge subgroup'' or ``split subgroup'' (\textbf{R3}). Analysis noise can be reduced by using a switch to hide non-subgroup units.

\subsection{Treatment Effect Validation View} \label{sec:validation-view}


The Validation View (\cref{fig:teaser}-C) has three parts: (1) The histogram depicts the treatment and control group units' propensity score distribution. 
(2) The treatment effect dot plot shows the distribution of ITE of sampled matched pairs in the treatment and control groups.
(3) The unit information table gives detailed treatment, covariates, and outcomes of matching units, enhancing users' trust (\textbf{R5}).


The histogram (\cref{fig:teaser}-C1) of unit distribution with varied propensity scores indicates covariate similarity across units. The propensity score is on the horizontal axis, and the number of units is on the vertical. The treatment and control groups had yellow and pink distributions. The overlapping distribution is brown. More overlap between these two groups' distributions means better covariate balancing, reducing effect estimation bias. The top of the histogram shows the number of units in the two groups and matched pairs.


The treatment effect dot plot (\cref{fig:teaser}-C2) helps to understand how effect is calculated using the propensity score matching. Each dot represents a matched pair of treatment and control units. A dot's horizontal position indicates the ITE. The average of ITE is represented as vertical lines, while confidence intervals are represented as gray rectangles at the bottom of the axis.
The unit information table (\cref{fig:teaser}-C3) provides details on matched units. The left columns represent unit ID, propensity score, treatment, and outcome. Other columns are covariates. Clicking on the dot in the dot plot highlights corresponding rows. Unit-based data can help users understand how treatment affects outcomes through particular examples.
\section{Evaluation}

In this section, we implement quantitative experiments to evaluate the causal subgroup discovery model. Two case studies and expert interviews further validate the usefulness of \sysname.

\subsection{Quantitative Experiments}

The quantitative experiment aims to assess the efficacy of the proposed model in identifying subgroups with significant treatment effects. Based on multiple synthetic datasets and real datasets, we compare the model with various baselines on different metrics.

\textbf{Datasets}. We employed synthetic datasets and real-world datasets. Following the settings in~\cite{athey2016recursive, Wu2023StableEO}, we sampled units under the assumption of unconfoundedness. Some covariates are categorical, and others are numerical with a normal distribution. We simulated non-random treatment assignment in observational data by creating a treatment variable $T$ determined by a Bernoulli distribution. We also produced the treatment effect $\text{TE}$ and the outcome $Y$, calculated from covariates and parameter vectors. Categorical covariates are converted to one-hot encoding for computation. For detailed descriptions of the generation process, please see \href{https://osf.io/jaqmf/?view_only=ac9575209945476b955bf829c85196e9}{OSF}. We also collected real-world dataset including Twins\footnote{\url{https://github.com/AMLab-Amsterdam/CEVAE/tree/master/datasets/TWINS}} and IHDP\footnote{\url{https://search.r-project.org/CRAN/refmans/bartcs/html/ihdp.html}}. The details of the datasets are shown in \cref{tab:data_sta}.

\begin{table}[tb]
\centering
\caption{
Dataset statistics for quantitative experiments.
}
{
\begin{tabular}{lcccc}
\hline
Dataset & \#Units & \#Categorical & \#Numerical  \\ \hline
Syn-1 & 3000 & 5 & 5 \\
Syn-2 & 3000 & 5 & 15 \\
Syn-3 & 4000 & 5 & 25 \\ 
Syn-4 & 4000 & 5 & 45 \\ 
Syn-5 & 4000 & 5 & 75 \\ 
Syn-6 & 4000 & 5 & 95 \\ 
Twins & 23968 & 3 & 46  \\
IHDP & 747 & 19 &  6 \\ \hline
\end{tabular}
}
\label{tab:data_sta}
\end{table}

\textbf{Baselines}. We compare the proposed model with two groups of algorithms. The first group is the popular HTE estimation algorithms: (1) Causal Tree (CT)~\cite{athey2016recursive}; (2) Causal Forest (CF)~\cite{wager2018estimation}; and (3) Causal Rule Ensemble (CRE)~\cite{Lee2020CausalRE}. The second group is the rule learning and subgroup discovery algorithms: (1) BRCG~\cite{Dash2018BooleanDR}; (2) Decision Tree (DT)~\cite{Breiman2017PointsOS}; (3) Pysubgroup (PYS)~\cite{lemmerich2018pysubgroup}. In the first group, CRE can explicitly obtain the antecedent and treatment effect of the subgroup. For CT and CF, it can be considered that the path from the root to the leaf nodes in the tree structure is the antecedent of the causal subgroup. The second group of methods can only get the correlation subgroups. In order to adapt to the causality setting, we add a post-processing step. CATE and variance are calculated on the data covered by each subgroup via \cref{eq:effect} and \cref{eq:variance}.

\textbf{Metrics}. We evaluate the quality of causal subgroups obtained from different perspectives. First, in order to evaluate the multi-objective optimization of treatment effect and outcome variance, it is proposed that (1) Precision(P) = (the true number of dominating subgroups)/(the number of subgroups in the front discovered by the method). Due to the lack of ground truth for subgroups belonging to the Pareto front, we collected subgroups in the front obtained by all methods and assumed that a subgroup is considered a true dominating subgroup if it is not dominated by any other subgroup. We also considered the interpretability of subgroups, including the metrics (2) \#Subgroups(S) = number of subgroups in front, (3) Avg\_len(L) = average length of antecedent(\ie number of covariates) used to describe the subgroups and (4) Coverage(C\%) = The average percent of units in a subgroup to the total number of units.


\textbf{Result analysis}. The experimental results are reported in \cref{tab:exp_result}. 
Our model has near-perfect precision (the bigger, the better), indicating that for ``Pareto fronts'' in other methods, our model always finds a dominant subgroup that is better in at least one objective. This is mainly due to the fact that we directly formalize and solve the constrained multi-objective optimization problem. Some methods use a two-stage approach to subgroup generation and selection, such as CT and CF, which partition the covariate space through trees and select the best subgroups. DT, on the other hand, only considers covariates and outcomes, ignoring treatment changes. Missing important subgroups in the initial stage may result in suboptimal results and reduced precision. We also uncover more diverse subgroups, averaging 15.4 subgroups distributed over the Pareto front, which is about 3 times that of other techniques. The small antecedent length and coverage of our method indicate fine-grained subgroups. Adjusting the minimal coverage limit can yield larger subgroups. As the number of covariates increases (10 to 100), our model remains stable, demonstrating that the efficient heuristic genetic algorithm can find satisfactory subgroups in high-dimensional covariate spaces, whereas other approaches or manual selection fail.

\begin{table*}[htb]
    \centering
    \caption{Quantitative metrics for different causal subgroups discovery methods. \textbf{Black}: best. \underline{Underline}: second best. (The experimental results of CRE and BRCG on Twins are missing due to code execution errors.)}
    \label{tab:exp_result}
\setlength\tabcolsep{2.3pt}
\begin{tabular}{c|cccc|cccc|cccc|cccc|cccc|cccc|cccc}
\toprule

Methods &\multicolumn{4}{c|}{Ours}&\multicolumn{4}{c|}{CT}&\multicolumn{4}{c|}{CF}&\multicolumn{4}{c|}{CRE}&\multicolumn{4}{c|}{DT}&\multicolumn{4}{c|}{PYS}&\multicolumn{4}{c}{BRCG} \\

\midrule

Metric & P & S & L & C & P & S & L & C & P & S & L & C & P & S & L & C & P & S & L & C & P & S & L & C & P & S & L & C \\
\midrule

Syn-1
& \bf{1.0} & 17 & 4.0 & 3.8 & 0.0 & 6 & 6.0 & 7.0 & 0.3 & 8 & 5.7 & 3.2 & \underline{0.5} & 2 & 2.3 & 30.2 & 0.0 & 8 & 3.8 & 9.1 & 0.0 & 3 & 1.4 & 13.7 & 0.0 & 2 & 3.0 & 12.5  \\
\midrule

Syn-2
& \bf{0.9} & 14 & 3.0 & 3.5 & \underline{0.3} & 3 & 4.3 & 12.8 & 0.0 & 7 & 4.5 & 3.2 & 0.1 & 8 & 2.3 & 28.7 & 0.1 & 7 & 4.0 & 8.6 & 0.0 & 3 & 1.2 & 16.9 & 0.0 & 3 & 3.0 & 11.8 \\
\midrule

Syn-3
& \bf{1.0} & 17 & 3.8 & 3.7 & \bf{1.0} & 3 & 6.4 & 5.8 & 0.0 & 6 & 5.2 & 3.1 & 0.0 & 3 & 2.1 & 32.7 & 0.0 & 4 & 4.0 & 9.9 & 0.0 & 5 & 1.0 & 20.1 & 0.0 & 3 & 3.2 & 9.1 \\
\midrule

Syn-4
& \bf{1.0} & 18 & 3.6 & 3.5 & 0.4 & 5 & 6.2 & 7.6 & \underline{0.5} & 2 & 4.4 & 2.7 & 0.2 & 5 & 1.7 & 32.2 & 0.2 & 5 & 4.0 & 8.3 & 0.0 & 8 & 1.0 & 20.1 & 0.0 & 1 & 10.0 & 35.4 \\
\midrule

Syn-5
& \underline{0.9} & 19 & 4.0 & 3.4 & 0.0 & 7 & 8.5 & 5.2 & 0.0 & 2 & 5.5 & 2.5 & \bf{1.0} & 1 & 1.7 & 35.7 & 0.0 & 4 & 4.0 & 12.5 & 0.0 & 3 & 1.0 & 20.0 & 0.0 & 1 & 8.0 & 33.5 \\
\midrule

Syn-6
& \bf{1.0} & 13 & 3.1 & 3.2 & 0.0 & 6 & 5.4 & 10.6 & 0.0 & 1 & 4.4 & 2.6 & 0.0 & 3 & 2.6 & 31.3 & 0.0 & 3 & 4.0 & 9.6 & 0.0 & 11 & 1.0 & 20.1 & 0.0 & 1 & 1.0 & 40.0 \\
\midrule


Twins
& \underline{0.8} & 15 & 4.0 & 4.0 & 0.2 & 5 & 5.1 & 9.7 & \bf{1.0} & 2 & 5.7 & 4.4 & / & / & / & / & 0.0  & 2 & 2.9 & 13.9 & 0.0 & 7 & 2.6 & 21.1 & / & / & / & / \\
\midrule

IHDP
& \bf{1.0} & 10 & 4.0 & 3.1 & \underline{0.1} & 7 & 4.1 & 7.5 & 0.0 & 11 & 3.1 & 11.9 & 0.0 & 3 & 1.5 & 28.8 & 0.0 & 3 & 2.6 & 18.1 & 0.0 & 2 & 2.8 & 19.4 & 0.0 & 2 & 2.5 & 24.2 \\
\midrule

Average & \bf{1.0} & 15.4 & 3.7 & 3.5 & \underline{0.3} & 5.3 & 5.8 & 8.3 & 0.2 & 4.9 & 4.8 & 4.2 & \underline{0.3} & 3.6 & 2.0 & 31.4 & 0.0 & 4.5 & 3.7 & 11.0 & 0.0 & 5.3 & 1.5 & 19.0 & 0.0 & 1.9 & 4.4 & 23.8 \\

\bottomrule
\end{tabular}
\end{table*}

\subsection{Case Studies}

We use two cases to demonstrate the analysis process of \sysname to explore causal heterogeneity from real datasets.

\subsubsection{Case 1: Default of Credit Card Clients}

\begin{figure}[tbp]
    \centering 
    \includegraphics[width=\columnwidth]{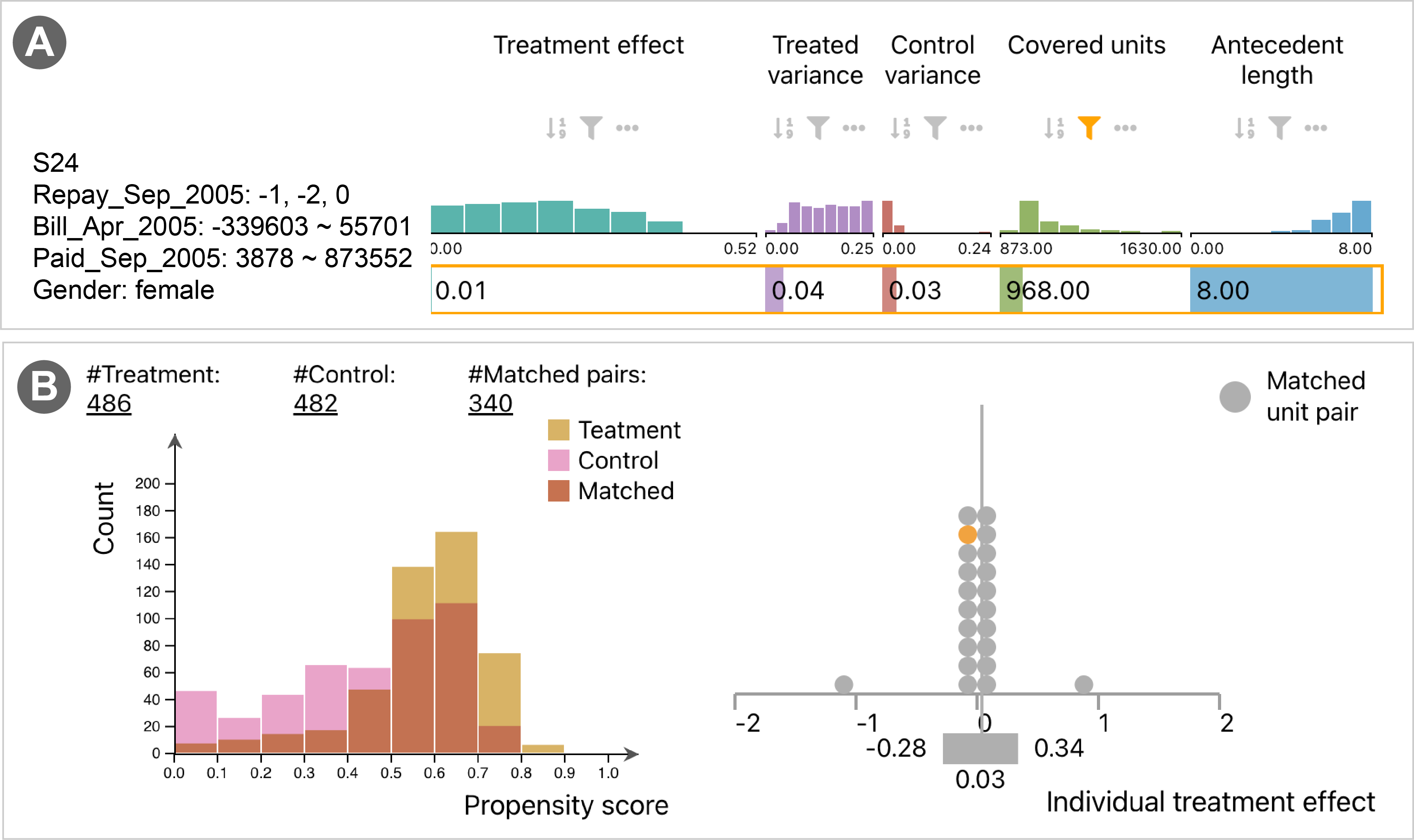}
    \caption{Descriptions of a subgroup with good credit in Case 1. (A) Explanation of treatment effects shows that the subgroup has low effect and variance. (B) A balanced histogram of propensity scores and most matched pairs have a zero ITE.}
    \label{fig:case1-2}
\end{figure}


This dataset contains behavioral data of about 30,000 credit card customers. The treatment is the credit limit (1 for greater than 200,000 and 0 for less), the outcome is default status (1 is defaulted, 0 is not), and the covariates include gender, education, marital status, age, and historical bill amounts \etc An account manager plans to raise credit limits to boost interest profits and transaction volume but avoids default risk by using \sysname to identify acceptable credit recipients.

He first loads the observational data on client behavior and sets the maximum subgroup antecedent length to 7 and the minimum coverage to 5\%. Afterward, the Causal Subgroup View (\cref{fig:teaser}-A1) displays the causal subgroups that the model automatically mines from the data (\textbf{R1}). He scrolls through the table to see descriptions of the covariates involved in the subgroup antecedents and clicks to see the details (\textbf{R2}). 
He found that two subgroups had similar distributions of covariates and were both of small size, so he clicked the ``Merge Subgroups'' button to generate a large one (\textbf{R3}).
He wanted to study risky subgroups and creditworthy subgroups, but checking each subgroup one by one was tedious, so he used the multi-attributes ranking function.
To find risky subgroups, he first set an inverted mapping from a variance value to the length of a bar, that is, a longer bar represents less variance. Then, he combined the treatment effects, the outcome variances of the treatment and the control group as objectives of subgroup identification and assigned weights of 6:2:2.
As shown in \cref{fig:teaser}-A3, after sorting in descending order of the combined column and filtering by the number of covered units greater than 950, S98 rose to the first place (\textbf{R4}), which includes past bill amounts, repayment records, \etc S98 has an effect of 0.4, a treatment group outcome variance of 0.24, and a control group outcome variance of 0.03.
He explained that these customers often overdue their repayments (the repay status indicates that the number of overdue months is up to 8), and their recent bill amounts are relatively high. Therefore, increasing their credit limit makes them more prone to defaults and a priority group to monitor.

To further verify the causal subgroup (\textbf{R5}), he went to the Treatment Effect Validation View (\cref{fig:teaser}-C). In S98, the treatment group contains 722 units, the control group contains 238 units, and the number of matched unit pairs reaches 208. The propensity score histogram proves that for each unit in the control group, a similar matching unit can be found in the treatment group almost all the time. In the dot plot of the ITE, the dots are mostly distributed in the middle or on the right side, indicating that increasing the credit limit has a positive effect on default. The detailed information in the table suggests that the outcome goes from 0 to 1 (\ie from non-default to default) in matched pairs of similar units (\eg units with id 2184 and 6246) after the treatment, which enhances his trust in the causal conclusion.

He next sought higher-credit subgroups for comparison. According to experience, these subgroups should have lower effects and variances, so he canceled the invert mapping of variances and sorted the combined columns in ascending order (\textbf{R4}). As shown in \cref{fig:case1-2}, the first subgroup, S24, caught his attention. It had an effect of 0.01, a treatment group outcome variance of 0.04, a control group outcome variance of 0.03, and a coverage of 968 people. Based on the antecedent description, he believed that this group had good repayment records in the past (the repay status is -1, -2, and 0) and had the habit of automatic repayment (the presence of a negative value of bill represents an automatic deposits into the credit card every month). The dot plot with dots concentrated at 0 also proves that increasing the limit hardly causes defaults.

\subsubsection{Case 2: Bank Marketing}

The dataset comes from a bank's marketing campaign, in which the treatment is the number of phone calls to customers (1 for more than two times, 0 otherwise), and the outcome is whether the customer makes a deposit (1 yes, 0 no), and the other 15 covariates consist of the customer's age, occupation, marital status, education level, recent contact data, and other socioeconomic indicators. An analyst wanted to identify the groups where increased contact could lead to deposits.

\begin{figure}[tbp]
    \centering 
    \includegraphics[width=\columnwidth]{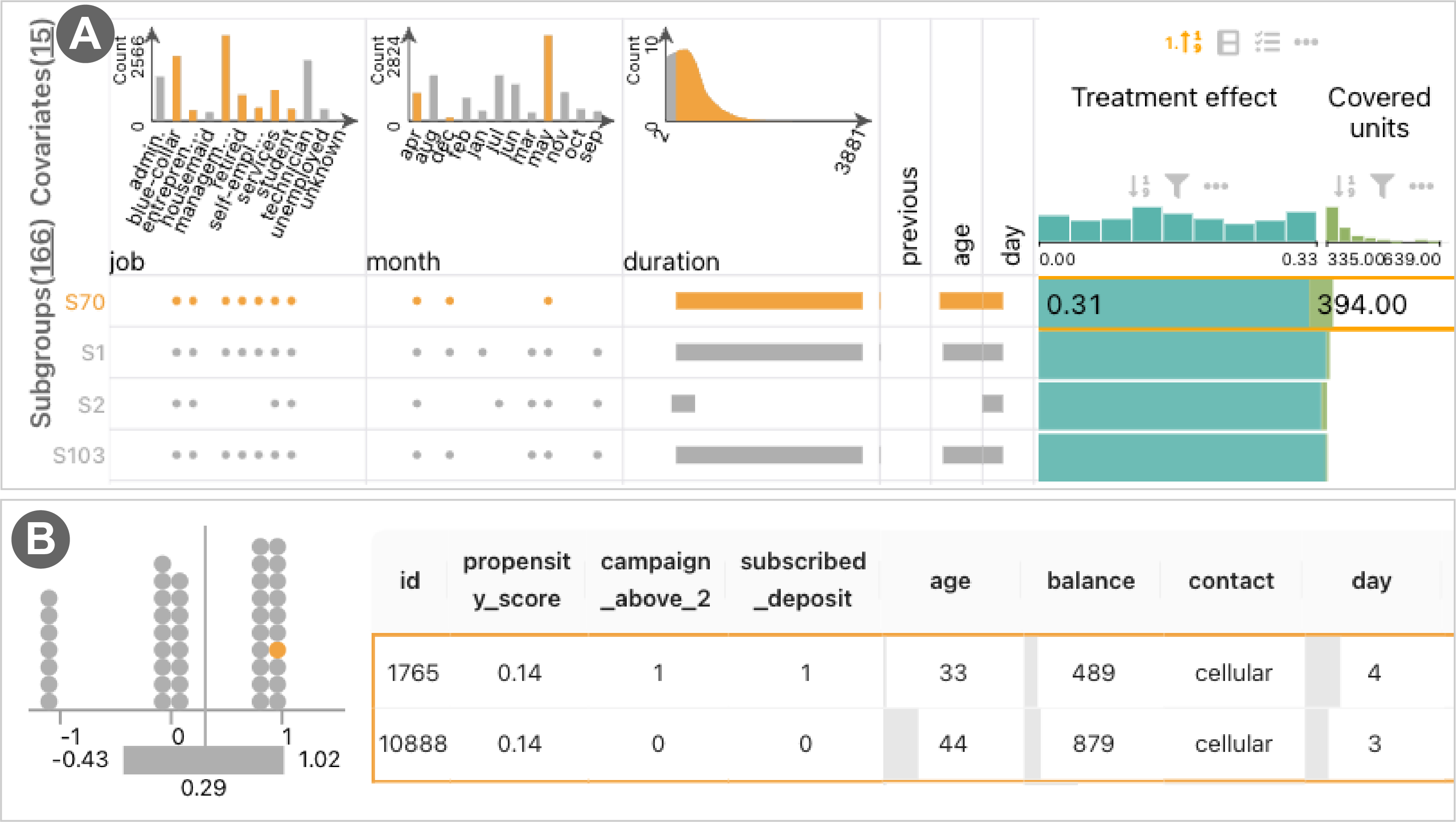}
    \caption{The subgroup identified in Case 2 as suitable for multiple contact to market deposits. (A) This subgroup is mostly people who have good jobs and showed interest in the last communication. They also have a large treatment effect and coverage rate. (B) The dot plot of ITE is skewed towards the strong side of the effect. Although it has a large variance, it is within the acceptable range of the user.}
    \label{fig:case2-1}
\end{figure}

With the Causal Subgroup View, she found that many subgroups have covariates of job, month, and duration, aligning with her domain knowledge, as there are differences in the financial capacity of clients across occupations and deposits have low and peak times of the year(\textbf{R1},\textbf{R2}). The length of the last contact also reflects the customer's wishes. Since telemarketing failure will not bring serious consequences, she placed greater emphasis on the effect strength and customer coverage (\textbf{R4}). Then, she combined the effect and coverage metrics, set the weights to 70\% and 30\% respectively, and sorted to get the subgroup S70 that suits her preference (\cref{fig:case2-1}-A). The subgroup is described as job: blue-collar, entrepreneur, management, retired, \etc; duration: 204-3881, age: 31- 95; contact day: 1-13. The corresponding effect is 0.31 with 394 covered units. She guessed that these middle-aged and elderly people had a sound financial foundation, and previous contacts also showed that they were interested. In addition, the beginning of the month is suitable for promoting deposits because wages are usually just paid.

To verify the reliability of the effect, she turned to the Treatment Effect Validation View. She recognized that almost all units in the treatment group had matching counterparts in the control group, ensuring the validity of the effect estimates. The dot plot revealed a high mean and a wide spread of dots, implying greater uncertainty (\textbf{R5}). Leveraging the detailed information provided by the table (\cref{fig:case2-1}-B), she found that an increase in the number of contacts indeed promotes deposits, so she decided to start by selecting those customers (\eg the user with id 10888) with fewer contacts in the matched pair with an individual treatment effect of 1 since they are more likely to purchase deposits after several more contacts.

\subsection{Expert Interviews}

We conducted informal interviews with four experts, two of whom (E1, E2) are data analysts at technology companies, E3 is a Ph.D. student studying causality, and E4 is a researcher at a university medical school who studies big data mining for medical data. They had not been involved in the previous design process.

\textbf{Procedure}. Each interview consisted of four steps, which are an introduction of background (5min), a demo of usage cases (15min), a think-aloud exploration (15min), and a talk for feedback collection (10min). Finally, we summarized comments, which includes reviews on on both the proposed model and the prototype system, and suggestions.

\textbf{Reviews on the proposed model}. Experts believe the causal subgroup discovery model can find subgroups with significant treatment effects compared to the population average. E1 said, ``When facing multi-dimensional observation data, I am usually left with time-consuming manual subgroup analysis, whereas this model gives me quick guidance on subgroups and helps to uncover unexpected conclusions.'' E3 praised that rules make subgroups easier to understand than clustering. Also, causal tree or forest-based approaches include numerous repeated covariate descriptions since many leaves share ancestor routes. Solving subgroups with multi-objective optimization is straightforward and yields better results. E4 stated that it would be more effective if they could freely combine different objectives and constraints.

\textbf{Reviews on the prototype system}. Experts confirmed that \sysname's visual design is easy to grasp. E1 and E2 thought the Causal Subgroup View was intuitive because they utilized similar Excel spreadsheets for work. E4 recommended dot plots for their clear explanation of the trade-off between bias and variation in treatment effects estimation. E3 said the multi-attribute ranking visualization is useful for decision-making situations where users have diverse preferences for several objectives, such as weighing revenues and costs when choosing an advertisement audience. E4 highlighted that adding and modifying subgroups helps incorporate domain knowledge and improves analytical freedom beyond model outputs.

\textbf{Suggestions}. Experts offered insightful advice on \sysname. E2's advice is to the function of merging subgroups with neighboring attribute values on the covariates, which can reduce the number of fragmented subgroups. Taking the advice, we enabled subgroup editing options, consisting of merging or splitting subgroups based on similarity and size, in the Covariate Projection View. E4 told us that manual editing of subgroups would be tedious if many predefined subgroups needed to be introduced. It would be nice if the Causal Subgroup View could automatically generate subgroups that have been recorded by experts by connecting with external knowledge bases. We plan to explore this feature by taking knowledge graphs as input.
\section{Discussion}

This section discusses the scalability, lessons learned, and the limitations and future work of \sysname.

\textbf{Scalability}. (1) The running efficiency of the proposed model is mainly affected by the number of subgroups $P$. The complexity in the subgroup generation stage is $O(Pd)$, where each subgroup could be described by at most $d$ covariates. The following three steps, checking the dominance relation in pairs, calculating the crowding distance of each subgroup, and sorting and selection, have complexities of $O(P^2)$, $O(P)$, and $O(P\log P)$, respectively. The overall time complexity is $O(Pd + P ^2)$. 
(2) For visualization, the covariate table may become crowded as the number of covariates increases; we moved the covariates that appear more frequently in the subgroup antecedents forward and added sliders and column folding functions to alleviate this. The dot plot in the Treatment Effect Validation View may appear to be overplotted due to an excessive number of matching units. We used sampling to reduce the number of dots and maintain the shape of the distribution. Other possible solutions are non-linear dot plots with adaptive dot sizes~\cite{8017644}.

\textbf{Lessons learned}. We gained valuable insights by developing \sysname. (1) Whether to analyze the heterogeneity of the causal graph structure or the HTE depends on the task. Causal discovery can identify many causal relationships among variables, such as gene regulatory networks. In the absence of prior knowledge of the causal structure, causal inference is a more general method, such as studying the treatment effect of policies on economic indicators. (2) Causal heterogeneity analysis benefits from multi-objective optimization. Our initial optimization target was treatment effect strength, but experts stressed that in actual applications, they would also evaluate many objectives, such as outcome variance, cost per unit, and return on investment (ROI). We then reduce multi-objectives optimization to single-objectives optimization by weighting (\eg $\tau + w_0 \sigma^2(0) + w_1 \sigma^2(1)$). However, experts said weights are sensitive and hard to modify, so the Pareto front of the MOO problem was eventually learned. We take effect and variances as objectives, but they can be expanded flexibly, which helps to meet diverse user preferences. (3) Explanations can support human-in-the-loop causal analysis, but too much information causes cognitive overload. Our initial prototype solution used a white-box causal subgroup discovery model based on decision trees, trading performance for interpretability. However, experts said they value the model's precision in real scenarios. If the model is bad, explaining it is pointless. Thus, we proposed an effective causal subgroup discovery model with a popular CATE estimator (IPW) and a heuristic search algorithm. We also provided post-hoc visualizations of subgroups and treatment effects to aid user interpretation.

\textbf{Limitations and future work}. Three limitations are observed in \sysname. First, subgroups support understanding data from the level of groups but may obscure individual uniqueness. For numerical attributes, units in the same subgroup may still have subtle covariate differences, and similar covariate values may belong to separate subgroups due to boundary divisions. Also, subgroup quality can be assessed based on external or internal similarity metrics. 
Second, our approach considers a single outcome. A treatment may have multiple outcomes. For example, a medicine may cause mortality, disease progression, or adverse events. A comprehensive evaluation of the treatment effectiveness should consider all possible outcomes. Future plans include combining causal inference approaches like mediation analysis~\cite{ten2012review} and multi-task learning~\cite{alaa2017bayesian} for multiple outcomes.
Third, we assume enough confounding variables in observational data. However, unobserved variables may distort effect estimates in some cases. For example, patients' diets may also affect treatment effects. Combining RCTs and observational studies~\cite{colnet2020causal} is a promising approach.

\section{Conclusion}

In this paper, we propose a visual analytics approach to support users in identifying, comparing, ranking, and validating subgroups with significant treatment effects in observational data. We first introduce interpretable rules to describe subgroups and then formalize the causal subgroup discovery into a constrained multi-objective optimization problem, whose corresponding Pareto optimal subgroups are efficiently solved by a heuristic genetic algorithm. A visual analysis system, \sysname, is developed to implement the proposed approach, which features a series of visualizations such as subgroups—covariate tables, multi-attribute rankings, and treatment effect explanations. Quantitative experiments prove that our model outperforms state-of-the-art methods in terms of precision and readability. Case studies and expert interviews demonstrate the usability of the system in helping users explore and interpret subgroups that satisfy different preferences.

\acknowledgments{%
	We would like to thank all the reviewers for their constructive comments. 
This work was supported by the National Natural Science Foundation of China (\# 62132017) and Zhejiang Provincial Natural Science Foundation of China (\# LD24F020011).%
}

\newpage
\bibliographystyle{src/abbrv-doi-hyperref}

\bibliography{main}

\begin{thebibliography}{10}

\bibitem{alaa2017bayesian}
A.~M. Alaa and M.~Van Der~Schaar.
\newblock Bayesian inference of individualized treatment effects using multi-task gaussian processes.
\newblock {\em Advances in Neural Information Processing Systems}, 30, 2017.

\bibitem{athey2016recursive}
S.~Athey and G.~Imbens.
\newblock Recursive partitioning for heterogeneous causal effects.
\newblock {\em Proc. NAS}, 113(27):7353--7360, 2016. \href{https://doi.org/10.1073/pnas.1510489113}
{doi: {{%
10\hspace{.1pt}\discretionary{.}{%
}{.}\hspace{.4pt}1073\discretionary{/}{%
}{/}pnas\hspace{.1pt}\discretionary{.}{%
}{.}\hspace{.4pt}1510489113}}}


\bibitem{athey2019estimating}
S.~Athey and S.~Wager.
\newblock Estimating treatment effects with causal forests: An application.
\newblock {\em Observational Studies}, 5(2):37--51, 2019. \href{https://doi.org/10.1353/obs.2019.0001}
{doi: {{%
10\hspace{.1pt}\discretionary{.}{%
}{.}\hspace{.4pt}1353\discretionary{/}{%
}{/}obs\hspace{.1pt}\discretionary{.}{%
}{.}\hspace{.4pt}2019\hspace{.1pt}\discretionary{.}{%
}{.}\hspace{.4pt}0001}}}


\bibitem{AtzmuellerSubgroupD}
M.~Atzmueller.
\newblock Subgroup discovery.
\newblock {\em WIREs Data Mining and Knowledge Discovery}, 5(1):35--49, 2015. \href{https://doi.org/10.1002/widm.1144}
{doi: {{%
10\hspace{.1pt}\discretionary{.}{%
}{.}\hspace{.4pt}1002\discretionary{/}{%
}{/}widm\hspace{.1pt}\discretionary{.}{%
}{.}\hspace{.4pt}1144}}}


\bibitem{Atzmller2009FastSD}
M.~Atzm{\"u}ller and F.~Lemmerich.
\newblock Fast subgroup discovery for continuous target concepts.
\newblock In {\em Proc. ISMIS}, p. 35–44, 2009. \href{https://doi.org/10.1007/978-3-642-04125-9_7}
{doi: {{%
10\hspace{.1pt}\discretionary{.}{%
}{.}\hspace{.4pt}1007\discretionary{/}{%
}{/}978\discretionary{%
}{-}{-}3\discretionary{%
}{-}{-}642\discretionary{%
}{-}{-}04125\discretionary{%
}{-}{-}9\_7}}}


\bibitem{Atzmller2006SDMapA}
M.~Atzm{\"u}ller and F.~Puppe.
\newblock Sd-map - a fast algorithm for exhaustive subgroup discovery.
\newblock In {\em Proc. ECML PKDD}, p. 6–17, 2006. \href{https://doi.org/10.1007/11871637_6}
{doi: {{%
10\hspace{.1pt}\discretionary{.}{%
}{.}\hspace{.4pt}1007\discretionary{/}{%
}{/}11871637\_6}}}


\bibitem{DBLP:journals/cgf/BaeHR17}
J.~Bae, T.~Helldin, and M.~Riveiro.
\newblock Understanding indirect causal relationships in node-link graphs.
\newblock {\em Computer Graphics Forum}, 36(3):411--421, 2017. \href{https://doi.org/10.1111/cgf.13198}
{doi: {{%
10\hspace{.1pt}\discretionary{.}{%
}{.}\hspace{.4pt}1111\discretionary{/}{%
}{/}cgf\hspace{.1pt}\discretionary{.}{%
}{.}\hspace{.4pt}13198}}}


\bibitem{DBLP:conf/grapp/BaeVRHF17}
J.~Bae, E.~Ventocilla, M.~Riveiro, T.~Helldin, and G.~Falkman.
\newblock Evaluating multi-attributes on cause and effect relationship visualization.
\newblock In {\em Proc. VISIGRAPP}, pp. 64--74, 2017. \href{https://doi.org/10.5220/0006102300640074}
{doi: {{%
10\hspace{.1pt}\discretionary{.}{%
}{.}\hspace{.4pt}5220\discretionary{/}{%
}{/}0006102300640074}}}


\bibitem{Lee2020CausalRE}
F.~J. Bargagli-Stoffi, R.~Cadei, K.~Lee, and F.~Dominici.
\newblock Causal rule ensemble: Interpretable discovery and inference of heterogeneous causal effects.
\newblock {\em arXiv preprint arXiv:2009.09036}, 2020. \href{https://doi.org/10.48550/arXiv.2009.09036}
{doi: {{%
10\hspace{.1pt}\discretionary{.}{%
}{.}\hspace{.4pt}48550\discretionary{/}{%
}{/}arXiv\hspace{.1pt}\discretionary{.}{%
}{.}\hspace{.4pt}2009\hspace{.1pt}\discretionary{.}{%
}{.}\hspace{.4pt}09036}}}


\bibitem{Blumberg2016CausalIF}
C.~J. Blumberg.
\newblock Causal inference for statistics, social, and biomedical sciences: An introduction.
\newblock {\em International Statistical Review}, 84(1):159--159, 2016. \href{https://doi.org/10.1111/insr.12170}
{doi: {{%
10\hspace{.1pt}\discretionary{.}{%
}{.}\hspace{.4pt}1111\discretionary{/}{%
}{/}insr\hspace{.1pt}\discretionary{.}{%
}{.}\hspace{.4pt}12170}}}


\bibitem{Breiman2017PointsOS}
L.~Breiman and R.~A. Olshen.
\newblock Points of significance: Classification and regression trees.
\newblock {\em Nature Methods}, 14:757--758, 2017. \href{https://doi.org/10.1038/nmeth.4370}
{doi: {{%
10\hspace{.1pt}\discretionary{.}{%
}{.}\hspace{.4pt}1038\discretionary{/}{%
}{/}nmeth\hspace{.1pt}\discretionary{.}{%
}{.}\hspace{.4pt}4370}}}


\bibitem{cai2011analysis}
T.~Cai, L.~Tian, P.~H. Wong, and L.~J. Wei.
\newblock Analysis of randomized comparative clinical trial data for personalized treatment selections.
\newblock {\em Biostatistics}, 12(2):270--282, 2011. \href{https://doi.org/10.1093/biostatistics/kxq060}
{doi: {{%
10\hspace{.1pt}\discretionary{.}{%
}{.}\hspace{.4pt}1093\discretionary{/}{%
}{/}biostatistics\discretionary{/}{%
}{/}kxq060}}}


\bibitem{Caliendo2005SomePG}
M.~Caliendo and S.~Kopeinig.
\newblock Some practical guidance for the implementation of propensity score matching.
\newblock {\em Journal of Economic Surveys}, 22(1):31--72, 2008. \href{https://doi.org/10.1111/j.1467-6419.2007.00527.x}
{doi: {{%
10\hspace{.1pt}\discretionary{.}{%
}{.}\hspace{.4pt}1111\discretionary{/}{%
}{/}j\hspace{.1pt}\discretionary{.}{%
}{.}\hspace{.4pt}1467\discretionary{%
}{-}{-}6419\hspace{.1pt}\discretionary{.}{%
}{.}\hspace{.4pt}2007\hspace{.1pt}\discretionary{.}{%
}{.}\hspace{.4pt}00527\hspace{.1pt}\discretionary{.}{%
}{.}\hspace{.4pt}x}}}


\bibitem{9229232}
F.~Cheng, Y.~Ming, and H.~Qu.
\newblock Dece: Decision explorer with counterfactual explanations for machine learning models.
\newblock {\em IEEE Transactions on Visualization and Computer Graphics}, 27(2):1438--1447, 2021. \href{https://doi.org/10.1109/TVCG.2020.3030342}
{doi: {{%
10\hspace{.1pt}\discretionary{.}{%
}{.}\hspace{.4pt}1109\discretionary{/}{%
}{/}TVCG\hspace{.1pt}\discretionary{.}{%
}{.}\hspace{.4pt}2020\hspace{.1pt}\discretionary{.}{%
}{.}\hspace{.4pt}3030342}}}


\bibitem{colnet2020causal}
B.~Colnet et~al.
\newblock Causal inference methods for combining randomized trials and observational studies: a review.
\newblock {\em arXiv preprint arXiv:2011.08047}, 2020. \href{https://doi.org/10.48550/arXiv.2011.08047}
{doi: {{%
10\hspace{.1pt}\discretionary{.}{%
}{.}\hspace{.4pt}48550\discretionary{/}{%
}{/}arXiv\hspace{.1pt}\discretionary{.}{%
}{.}\hspace{.4pt}2011\hspace{.1pt}\discretionary{.}{%
}{.}\hspace{.4pt}08047}}}


\bibitem{Dash2018BooleanDR}
S.~Dash, O.~G\"{u}nl\"{u}k, and D.~Wei.
\newblock Boolean decision rules via column generation.
\newblock In {\em Proc. NeurIPS}, 2018. \href{https://doi.org/10.48550/arXiv.1805.09901}
{doi: {{%
10\hspace{.1pt}\discretionary{.}{%
}{.}\hspace{.4pt}48550\discretionary{/}{%
}{/}arXiv\hspace{.1pt}\discretionary{.}{%
}{.}\hspace{.4pt}1805\hspace{.1pt}\discretionary{.}{%
}{.}\hspace{.4pt}09901}}}


\bibitem{996017}
K.~Deb, A.~Pratap, S.~Agarwal, and T.~Meyarivan.
\newblock A fast and elitist multiobjective genetic algorithm: Nsga-ii.
\newblock {\em IEEE Transactions on Evolutionary Computation}, 6(2):182--197, 2002. \href{https://doi.org/10.1109/4235.996017}
{doi: {{%
10\hspace{.1pt}\discretionary{.}{%
}{.}\hspace{.4pt}1109\discretionary{/}{%
}{/}4235\hspace{.1pt}\discretionary{.}{%
}{.}\hspace{.4pt}996017}}}


\bibitem{Jess2007EvolutionaryFR}
M.~J. del Jes{\'u}s, P.~Gonz{\'a}lez, F.~Herrera, and M.~Mesonero.
\newblock Evolutionary fuzzy rule induction process for subgroup discovery: A case study in marketing.
\newblock {\em IEEE Transactions on Fuzzy Systems}, 15(4):578--592, 2007. \href{https://doi.org/10.1109/TFUZZ.2006.890662}
{doi: {{%
10\hspace{.1pt}\discretionary{.}{%
}{.}\hspace{.4pt}1109\discretionary{/}{%
}{/}TFUZZ\hspace{.1pt}\discretionary{.}{%
}{.}\hspace{.4pt}2006\hspace{.1pt}\discretionary{.}{%
}{.}\hspace{.4pt}890662}}}


\bibitem{DBLP:journals/tvcg/DengWXBZXCW22}
Z.~Deng, D.~Weng, X.~Xie, J.~Bao, Y.~Zheng, M.~Xu, W.~Chen, and Y.~Wu.
\newblock Compass: Towards better causal analysis of urban time series.
\newblock {\em IEEE Transactions on Visualization and Computer Graphics}, 28(1):1051--1061, 2022. \href{https://doi.org/10.1109/TVCG.2021.3114875}
{doi: {{%
10\hspace{.1pt}\discretionary{.}{%
}{.}\hspace{.4pt}1109\discretionary{/}{%
}{/}TVCG\hspace{.1pt}\discretionary{.}{%
}{.}\hspace{.4pt}2021\hspace{.1pt}\discretionary{.}{%
}{.}\hspace{.4pt}3114875}}}


\bibitem{8464305}
D.~Dingen et~al.
\newblock Regressionexplorer: Interactive exploration of logistic regression models with subgroup analysis.
\newblock {\em IEEE Transactions on Visualization and Computer Graphics}, 25(1):246--255, 2019. \href{https://doi.org/10.1109/TVCG.2018.2865043}
{doi: {{%
10\hspace{.1pt}\discretionary{.}{%
}{.}\hspace{.4pt}1109\discretionary{/}{%
}{/}TVCG\hspace{.1pt}\discretionary{.}{%
}{.}\hspace{.4pt}2018\hspace{.1pt}\discretionary{.}{%
}{.}\hspace{.4pt}2865043}}}


\bibitem{doi:10.1177/1473871619878085}
K.~Furmanova et~al.
\newblock Taggle: Combining overview and details in tabular data visualizations.
\newblock {\em Information Visualization}, 19(2):114--136, 2020. \href{https://doi.org/10.1177/1473871619878085}
{doi: {{%
10\hspace{.1pt}\discretionary{.}{%
}{.}\hspace{.4pt}1177\discretionary{/}{%
}{/}1473871619878085}}}


\bibitem{Gamberger2002ExpertGuidedSD}
D.~Gamberger and N.~Lavrac.
\newblock Expert-guided subgroup discovery: Methodology and application.
\newblock {\em Journal of Artificial Intelligence Research}, 17:501--527, 2002. \href{https://doi.org/10.1613/jair.1089}
{doi: {{%
10\hspace{.1pt}\discretionary{.}{%
}{.}\hspace{.4pt}1613\discretionary{/}{%
}{/}jair\hspace{.1pt}\discretionary{.}{%
}{.}\hspace{.4pt}1089}}}


\bibitem{gangl2010causal}
M.~Gangl.
\newblock Causal inference in sociological research.
\newblock {\em Annual review of sociology}, 36(1):21--47, 2010. \href{https://doi.org/10.1146/annurev.soc.012809.102702}
{doi: {{%
10\hspace{.1pt}\discretionary{.}{%
}{.}\hspace{.4pt}1146\discretionary{/}{%
}{/}annurev\hspace{.1pt}\discretionary{.}{%
}{.}\hspace{.4pt}soc\hspace{.1pt}\discretionary{.}{%
}{.}\hspace{.4pt}012809\hspace{.1pt}\discretionary{.}{%
}{.}\hspace{.4pt}102702}}}


\bibitem{6634146}
S.~Gratzl, A.~Lex, N.~Gehlenborg, H.~Pfister, and M.~Streit.
\newblock Lineup: Visual analysis of multi-attribute rankings.
\newblock {\em IEEE Transactions on Visualization and Computer Graphics}, 19(12):2277--2286, 2013. \href{https://doi.org/10.1109/TVCG.2013.173}
{doi: {{%
10\hspace{.1pt}\discretionary{.}{%
}{.}\hspace{.4pt}1109\discretionary{/}{%
}{/}TVCG\hspace{.1pt}\discretionary{.}{%
}{.}\hspace{.4pt}2013\hspace{.1pt}\discretionary{.}{%
}{.}\hspace{.4pt}173}}}


\bibitem{cobalt}
N.~Greifer.
\newblock {\em cobalt: Covariate Balance Tables and Plots}, 2024.
\newblock R package version 4.5.4, https://github.com/ngreifer/cobalt.

\bibitem{Grosskreutz2009OnSD}
H.~Grosskreutz and S.~R{\"u}ping.
\newblock On subgroup discovery in numerical domains.
\newblock {\em Data Mining and Knowledge Discovery}, 19:210--226, 2009. \href{https://doi.org/10.1007/s10618-009-0136-3}
{doi: {{%
10\hspace{.1pt}\discretionary{.}{%
}{.}\hspace{.4pt}1007\discretionary{/}{%
}{/}s10618\discretionary{%
}{-}{-}009\discretionary{%
}{-}{-}0136\discretionary{%
}{-}{-}3}}}


\bibitem{Grosskreutz2008TightOE}
H.~Grosskreutz, S.~R{\"u}ping, and S.~Wrobel.
\newblock Tight optimistic estimates for fast subgroup discovery.
\newblock In {\em Proc. ECML PKDD}, pp. 440--456, 2008. \href{https://doi.org/10.1007/978-3-540-87479-9_47}
{doi: {{%
10\hspace{.1pt}\discretionary{.}{%
}{.}\hspace{.4pt}1007\discretionary{/}{%
}{/}978\discretionary{%
}{-}{-}3\discretionary{%
}{-}{-}540\discretionary{%
}{-}{-}87479\discretionary{%
}{-}{-}9\_47}}}


\bibitem{9623285}
G.~Guo, M.~Glenski, Z.~Shaw, E.~Saldanha, A.~Endert, S.~Volkova, and D.~Arendt.
\newblock Vaine: Visualization and ai for natural experiments.
\newblock In {\em Proc. VIS}, pp. 21--25, 2021. \href{https://doi.org/10.1109/VIS49827.2021.9623285}
{doi: {{%
10\hspace{.1pt}\discretionary{.}{%
}{.}\hspace{.4pt}1109\discretionary{/}{%
}{/}VIS49827\hspace{.1pt}\discretionary{.}{%
}{.}\hspace{.4pt}2021\hspace{.1pt}\discretionary{.}{%
}{.}\hspace{.4pt}9623285}}}


\bibitem{10.1145/3544548.3581236}
G.~Guo, E.~Karavani, A.~Endert, and B.~C. Kwon.
\newblock Causalvis: Visualizations for causal inference.
\newblock In {\em Proc. CHI}, CHI '23,  article no. 462,  20 pages, 2023. \href{https://doi.org/10.1145/3544548.3581236}
{doi: {{%
10\hspace{.1pt}\discretionary{.}{%
}{.}\hspace{.4pt}1145\discretionary{/}{%
}{/}3544548\hspace{.1pt}\discretionary{.}{%
}{.}\hspace{.4pt}3581236}}}


\bibitem{10.1145/3397269}
R.~Guo, L.~Cheng, J.~Li, P.~R. Hahn, and H.~Liu.
\newblock A survey of learning causality with data: Problems and methods.
\newblock {\em ACM Computing Surveys}, 53(4),  article no. 75,  37 pages, jul 2020. \href{https://doi.org/10.1145/3397269}
{doi: {{%
10\hspace{.1pt}\discretionary{.}{%
}{.}\hspace{.4pt}1145\discretionary{/}{%
}{/}3397269}}}


\bibitem{Hariton2018rctgold}
E.~Hariton and J.~J. Locascio.
\newblock Randomised controlled trials - the gold standard for effectiveness research: Study design: randomised controlled trials.
\newblock {\em BJOG}, 125(13):1716, June 2018.

\bibitem{Herrera2011AnOO}
F.~Herrera, C.~J. Carmona, P.~Gonz{\'a}lez, and M.~J. del Jesus.
\newblock An overview on subgroup discovery: foundations and applications.
\newblock {\em Knowledge and Information Systems}, 29:495--525, 2011. \href{https://doi.org/10.1007/s10115-010-0356-2}
{doi: {{%
10\hspace{.1pt}\discretionary{.}{%
}{.}\hspace{.4pt}1007\discretionary{/}{%
}{/}s10115\discretionary{%
}{-}{-}010\discretionary{%
}{-}{-}0356\discretionary{%
}{-}{-}2}}}


\bibitem{hill2011bayesian}
J.~L. Hill.
\newblock Bayesian nonparametric modeling for causal inference.
\newblock {\em Journal of Computational and Graphical Statistics}, 20(1):217--240, 2011. \href{https://doi.org/10.1198/jcgs.2010.08162}
{doi: {{%
10\hspace{.1pt}\discretionary{.}{%
}{.}\hspace{.4pt}1198\discretionary{/}{%
}{/}jcgs\hspace{.1pt}\discretionary{.}{%
}{.}\hspace{.4pt}2010\hspace{.1pt}\discretionary{.}{%
}{.}\hspace{.4pt}08162}}}


\bibitem{Hirano2000EfficientEO}
K.~Hirano, G.~Imbens, and G.~Ridder.
\newblock Efficient estimation of average treatment effects using the estimated propensity score.
\newblock {\em Econometrica}, 71(4):1161--1189, 2000. \href{https://doi.org/10.1111/1468-0262.00442}
{doi: {{%
10\hspace{.1pt}\discretionary{.}{%
}{.}\hspace{.4pt}1111\discretionary{/}{%
}{/}1468\discretionary{%
}{-}{-}0262\hspace{.1pt}\discretionary{.}{%
}{.}\hspace{.4pt}00442}}}


\bibitem{9507307}
M.~N. Hoque and K.~Mueller.
\newblock Outcome-explorer: A causality guided interactive visual interface for interpretable algorithmic decision making.
\newblock {\em IEEE Transactions on Visualization and Computer Graphics}, 28(12):4728--4740, 2022. \href{https://doi.org/10.1109/TVCG.2021.3102051}
{doi: {{%
10\hspace{.1pt}\discretionary{.}{%
}{.}\hspace{.4pt}1109\discretionary{/}{%
}{/}TVCG\hspace{.1pt}\discretionary{.}{%
}{.}\hspace{.4pt}2021\hspace{.1pt}\discretionary{.}{%
}{.}\hspace{.4pt}3102051}}}


\bibitem{10.1162/003465304323023651}
G.~W. Imbens.
\newblock {Nonparametric Estimation of Average Treatment Effects Under Exogeneity: A Review}.
\newblock {\em The Review of Economics and Statistics}, 86(1):4--29, 2004. \href{https://doi.org/10.1162/003465304323023651}
{doi: {{%
10\hspace{.1pt}\discretionary{.}{%
}{.}\hspace{.4pt}1162\discretionary{/}{%
}{/}003465304323023651}}}


\bibitem{DBLP:journals/tvcg/JinGCWGC21}
Z.~Jin, S.~Guo, N.~Chen, D.~Weiskopf, D.~Gotz, and N.~Cao.
\newblock Visual causality analysis of event sequence data.
\newblock {\em IEEE Transactions on Visualization and Computer Graphics}, 27(2):1343--1352, 2021. \href{https://doi.org/10.1109/TVCG.2020.3030465}
{doi: {{%
10\hspace{.1pt}\discretionary{.}{%
}{.}\hspace{.4pt}1109\discretionary{/}{%
}{/}TVCG\hspace{.1pt}\discretionary{.}{%
}{.}\hspace{.4pt}2020\hspace{.1pt}\discretionary{.}{%
}{.}\hspace{.4pt}3030465}}}


\bibitem{johansson2016learning}
F.~Johansson, U.~Shalit, and D.~Sontag.
\newblock Learning representations for counterfactual inference.
\newblock In {\em Proc. ICML}, vol.~48, pp. 3020--3029, 2016. \href{https://doi.org/10.48550/arXiv.1605.03661}
{doi: {{%
10\hspace{.1pt}\discretionary{.}{%
}{.}\hspace{.4pt}48550\discretionary{/}{%
}{/}arXiv\hspace{.1pt}\discretionary{.}{%
}{.}\hspace{.4pt}1605\hspace{.1pt}\discretionary{.}{%
}{.}\hspace{.4pt}03661}}}


\bibitem{DBLP:journals/tvcg/KadabaIL07}
N.~R. Kadaba, P.~Irani, and J.~Leboe.
\newblock Visualizing causal semantics using animations.
\newblock {\em IEEE Transactions on Visualization and Computer Graphics}, 13(6):1254--1261, 2007. \href{https://doi.org/10.1109/TVCG.2007.70528}
{doi: {{%
10\hspace{.1pt}\discretionary{.}{%
}{.}\hspace{.4pt}1109\discretionary{/}{%
}{/}TVCG\hspace{.1pt}\discretionary{.}{%
}{.}\hspace{.4pt}2007\hspace{.1pt}\discretionary{.}{%
}{.}\hspace{.4pt}70528}}}


\bibitem{DBLP:journals/tvcg/KaleWH22}
A.~Kale, Y.~Wu, and J.~Hullman.
\newblock Causal support: Modeling causal inferences with visualizations.
\newblock {\em IEEE Transactions on Visualization and Computer Graphics}, 28(1):1150--1160, 2022. \href{https://doi.org/10.1109/TVCG.2021.3114824}
{doi: {{%
10\hspace{.1pt}\discretionary{.}{%
}{.}\hspace{.4pt}1109\discretionary{/}{%
}{/}TVCG\hspace{.1pt}\discretionary{.}{%
}{.}\hspace{.4pt}2021\hspace{.1pt}\discretionary{.}{%
}{.}\hspace{.4pt}3114824}}}


\bibitem{kruskal1964nonmetric}
J.~B. Kruskal.
\newblock Nonmetric multidimensional scaling: a numerical method.
\newblock {\em Psychometrika}, 29(2):115--129, 1964.

\bibitem{9973226}
B.~C. Kwon et~al.
\newblock Rmexplorer: A visual analytics approach to explore the performance and the fairness of disease risk models on population subgroups.
\newblock In {\em Proc. VIS}, pp. 50--54, 2022. \href{https://doi.org/10.1109/VIS54862.2022.00019}
{doi: {{%
10\hspace{.1pt}\discretionary{.}{%
}{.}\hspace{.4pt}1109\discretionary{/}{%
}{/}VIS54862\hspace{.1pt}\discretionary{.}{%
}{.}\hspace{.4pt}2022\hspace{.1pt}\discretionary{.}{%
}{.}\hspace{.4pt}00019}}}


\bibitem{Lavra2004SubgroupDW}
N.~Lavrač, B.~Kavšek, P.~A. Flach, and L.~Todorovski.
\newblock Subgroup discovery with cn2-sd.
\newblock {\em The Journal of Machine Learning Research}, 5:153--188, 2004.

\bibitem{lemmerich2018pysubgroup}
F.~Lemmerich and M.~Becker.
\newblock pysubgroup: Easy-to-use subgroup discovery in python.
\newblock In {\em Proc. ECML PKDD}, pp. 658--662, 2018. \href{https://doi.org/10.1007/978-3-030-10997-4_46}
{doi: {{%
10\hspace{.1pt}\discretionary{.}{%
}{.}\hspace{.4pt}1007\discretionary{/}{%
}{/}978\discretionary{%
}{-}{-}3\discretionary{%
}{-}{-}030\discretionary{%
}{-}{-}10997\discretionary{%
}{-}{-}4\_46}}}


\bibitem{6876017}
A.~Lex, N.~Gehlenborg, H.~Strobelt, R.~Vuillemot, and H.~Pfister.
\newblock Upset: Visualization of intersecting sets.
\newblock {\em IEEE Transactions on Visualization and Computer Graphics}, 20(12):1983--1992, 2014. \href{https://doi.org/10.1109/TVCG.2014.2346248}
{doi: {{%
10\hspace{.1pt}\discretionary{.}{%
}{.}\hspace{.4pt}1109\discretionary{/}{%
}{/}TVCG\hspace{.1pt}\discretionary{.}{%
}{.}\hspace{.4pt}2014\hspace{.1pt}\discretionary{.}{%
}{.}\hspace{.4pt}2346248}}}


\bibitem{Li2015FromOS}
J.~Li et~al.
\newblock From observational studies to causal rule mining.
\newblock {\em ACM Transactions on Intelligent Systems and Technology}, 7(2):1--27, 2015. \href{https://doi.org/10.1145/2746410}
{doi: {{%
10\hspace{.1pt}\discretionary{.}{%
}{.}\hspace{.4pt}1145\discretionary{/}{%
}{/}2746410}}}


\bibitem{10443294}
J.~Li, C.~Lai, and X.~Yuan.
\newblock Subspace-map: Interactive visual analysis for subspace data with a map metaphor.
\newblock {\em IEEE Transactions on Visualization and Computer Graphics}, pp. 1--15, 2024. \href{https://doi.org/10.1109/TVCG.2024.3368094}
{doi: {{%
10\hspace{.1pt}\discretionary{.}{%
}{.}\hspace{.4pt}1109\discretionary{/}{%
}{/}TVCG\hspace{.1pt}\discretionary{.}{%
}{.}\hspace{.4pt}2024\hspace{.1pt}\discretionary{.}{%
}{.}\hspace{.4pt}3368094}}}


\bibitem{peters2017elements}
J.~Peters, D.~Janzing, and B.~Sch{\"o}lkopf.
\newblock {\em Elements of causal inference: foundations and learning algorithms}.
\newblock The MIT Press, 2017. \href{https://doi.org/10.1080/00949655.2018.1505197}
{doi: {{%
10\hspace{.1pt}\discretionary{.}{%
}{.}\hspace{.4pt}1080\discretionary{/}{%
}{/}00949655\hspace{.1pt}\discretionary{.}{%
}{.}\hspace{.4pt}2018\hspace{.1pt}\discretionary{.}{%
}{.}\hspace{.4pt}1505197}}}


\bibitem{8017644}
N.~Rodrigues and D.~Weiskopf.
\newblock Nonlinear dot plots.
\newblock {\em IEEE Transactions on Visualization and Computer Graphics}, 24(1):616--625, 2018. \href{https://doi.org/10.1109/TVCG.2017.2744018}
{doi: {{%
10\hspace{.1pt}\discretionary{.}{%
}{.}\hspace{.4pt}1109\discretionary{/}{%
}{/}TVCG\hspace{.1pt}\discretionary{.}{%
}{.}\hspace{.4pt}2017\hspace{.1pt}\discretionary{.}{%
}{.}\hspace{.4pt}2744018}}}


\bibitem{rothman2005causation}
K.~J. Rothman and S.~Greenland.
\newblock Causation and causal inference in epidemiology.
\newblock {\em American journal of public health}, 95(S1):S144--S150, 2005. \href{https://doi.org/10.2105/AJPH.2004.059204}
{doi: {{%
10\hspace{.1pt}\discretionary{.}{%
}{.}\hspace{.4pt}2105\discretionary{/}{%
}{/}AJPH\hspace{.1pt}\discretionary{.}{%
}{.}\hspace{.4pt}2004\hspace{.1pt}\discretionary{.}{%
}{.}\hspace{.4pt}059204}}}


\bibitem{10.1093/oxfordhb/9780199286546.003.0011}
J.~Sekhon.
\newblock {271 The Neyman— Rubin Model of Causal Inference and Estimation Via Matching Methods}.
\newblock In {\em {The Oxford Handbook of Political Methodology}}. Oxford University Press, 08 2008. \href{https://doi.org/10.1093/oxfordhb/9780199286546.003.0011}
{doi: {{%
10\hspace{.1pt}\discretionary{.}{%
}{.}\hspace{.4pt}1093\discretionary{/}{%
}{/}oxfordhb\discretionary{/}{%
}{/}9780199286546\hspace{.1pt}\discretionary{.}{%
}{.}\hspace{.4pt}003\hspace{.1pt}\discretionary{.}{%
}{.}\hspace{.4pt}0011}}}


\bibitem{Shimoni2019AnET}
Y.~Shimoni et~al.
\newblock An evaluation toolkit to guide model selection and cohort definition in causal inference.
\newblock {\em ArXiv}, abs/1906.00442, 2019.

\bibitem{ten2012review}
T.~R. Ten~Have and M.~M. Joffe.
\newblock A review of causal estimation of effects in mediation analyses.
\newblock {\em Statistical Methods in Medical Research}, 21(1):77--107, 2012.

\bibitem{10292663}
X.~Teng, Y.~Ahn, and Y.-R. Lin.
\newblock Vispur: Visual aids for identifying and interpreting spurious associations in data-driven decisions.
\newblock {\em IEEE Transactions on Visualization and Computer Graphics}, 30(1):219--229, 2024. \href{https://doi.org/10.1109/TVCG.2023.3326587}
{doi: {{%
10\hspace{.1pt}\discretionary{.}{%
}{.}\hspace{.4pt}1109\discretionary{/}{%
}{/}TVCG\hspace{.1pt}\discretionary{.}{%
}{.}\hspace{.4pt}2023\hspace{.1pt}\discretionary{.}{%
}{.}\hspace{.4pt}3326587}}}


\bibitem{Leeuwen2012DiverseSS}
M.~van Leeuwen and A.~J. Knobbe.
\newblock Diverse subgroup set discovery.
\newblock {\em Data Mining and Knowledge Discovery}, 25:208--242, 2012. \href{https://doi.org/10.1007/s10618-012-0273-y}
{doi: {{%
10\hspace{.1pt}\discretionary{.}{%
}{.}\hspace{.4pt}1007\discretionary{/}{%
}{/}s10618\discretionary{%
}{-}{-}012\discretionary{%
}{-}{-}0273\discretionary{%
}{-}{-}y}}}


\bibitem{varian2016causal}
H.~R. Varian.
\newblock Causal inference in economics and marketing.
\newblock {\em Proc. NAS}, 113(27):7310--7315, 2016. \href{https://doi.org/10.1073/pnas.1510479113}
{doi: {{%
10\hspace{.1pt}\discretionary{.}{%
}{.}\hspace{.4pt}1073\discretionary{/}{%
}{/}pnas\hspace{.1pt}\discretionary{.}{%
}{.}\hspace{.4pt}1510479113}}}


\bibitem{wager2018estimation}
S.~Wager and S.~Athey.
\newblock Estimation and inference of heterogeneous treatment effects using random forests.
\newblock {\em Journal of the American Statistical Association}, 113(523):1228--1242, 2018. \href{https://doi.org/10.1080/01621459.2017.1319839}
{doi: {{%
10\hspace{.1pt}\discretionary{.}{%
}{.}\hspace{.4pt}1080\discretionary{/}{%
}{/}01621459\hspace{.1pt}\discretionary{.}{%
}{.}\hspace{.4pt}2017\hspace{.1pt}\discretionary{.}{%
}{.}\hspace{.4pt}1319839}}}


\bibitem{DBLP:journals/tvcg/WangM16}
J.~Wang and K.~Mueller.
\newblock The visual causality analyst: An interactive interface for causal reasoning.
\newblock {\em IEEE Transactions on Visualization and Computer Graphics}, 22(1):230--239, 2016. \href{https://doi.org/10.1109/TVCG.2015.2467931}
{doi: {{%
10\hspace{.1pt}\discretionary{.}{%
}{.}\hspace{.4pt}1109\discretionary{/}{%
}{/}TVCG\hspace{.1pt}\discretionary{.}{%
}{.}\hspace{.4pt}2015\hspace{.1pt}\discretionary{.}{%
}{.}\hspace{.4pt}2467931}}}


\bibitem{DBLP:conf/ieeevast/WangM17}
J.~Wang and K.~Mueller.
\newblock Visual causality analysis made practical.
\newblock In {\em Proc. VAST}, pp. 151--161, 2017. \href{https://doi.org/10.1109/VAST.2017.8585647}
{doi: {{%
10\hspace{.1pt}\discretionary{.}{%
}{.}\hspace{.4pt}1109\discretionary{/}{%
}{/}VAST\hspace{.1pt}\discretionary{.}{%
}{.}\hspace{.4pt}2017\hspace{.1pt}\discretionary{.}{%
}{.}\hspace{.4pt}8585647}}}


\bibitem{wang2022domino}
J.~Wang and K.~Mueller.
\newblock Domino: Visual causal reasoning with time-dependent phenomena.
\newblock {\em IEEE Transactions on Visualization and Computer Graphics}, 2022. \href{https://doi.org/10.1109/tvcg.2022.3207929}
{doi: {{%
10\hspace{.1pt}\discretionary{.}{%
}{.}\hspace{.4pt}1109\discretionary{/}{%
}{/}tvcg\hspace{.1pt}\discretionary{.}{%
}{.}\hspace{.4pt}2022\hspace{.1pt}\discretionary{.}{%
}{.}\hspace{.4pt}3207929}}}


\bibitem{8456575}
D.~Weng, R.~Chen, Z.~Deng, F.~Wu, J.~Chen, and Y.~Wu.
\newblock Srvis: Towards better spatial integration in ranking visualization.
\newblock {\em IEEE Transactions on Visualization and Computer Graphics}, 25(1):459--469, 2019. \href{https://doi.org/10.1109/TVCG.2018.2865126}
{doi: {{%
10\hspace{.1pt}\discretionary{.}{%
}{.}\hspace{.4pt}1109\discretionary{/}{%
}{/}TVCG\hspace{.1pt}\discretionary{.}{%
}{.}\hspace{.4pt}2018\hspace{.1pt}\discretionary{.}{%
}{.}\hspace{.4pt}2865126}}}


\bibitem{Wrobel1997AnAF}
S.~Wrobel.
\newblock An algorithm for multi-relational discovery of subgroups.
\newblock In {\em Proc. ECML PKDD}, p. 78–87, 1997. \href{https://doi.org/10.1007/3-540-63223-9_108}
{doi: {{%
10\hspace{.1pt}\discretionary{.}{%
}{.}\hspace{.4pt}1007\discretionary{/}{%
}{/}3\discretionary{%
}{-}{-}540\discretionary{%
}{-}{-}63223\discretionary{%
}{-}{-}9\_108}}}


\bibitem{Wu2023StableEO}
A.~Wu, K.~Kuang, R.~Xiong, B.~Li, and F.~Wu.
\newblock Stable estimation of heterogeneous treatment effects.
\newblock In {\em Proc. ICML}, 2023.

\bibitem{8017645}
J.~Xia, F.~Ye, W.~Chen, Y.~Wang, W.~Chen, Y.~Ma, and A.~K. Tung.
\newblock Ldsscanner: Exploratory analysis of low-dimensional structures in high-dimensional datasets.
\newblock {\em IEEE Transactions on Visualization and Computer Graphics}, 24(1):236--245, 2018. \href{https://doi.org/10.1109/TVCG.2017.2744098}
{doi: {{%
10\hspace{.1pt}\discretionary{.}{%
}{.}\hspace{.4pt}1109\discretionary{/}{%
}{/}TVCG\hspace{.1pt}\discretionary{.}{%
}{.}\hspace{.4pt}2017\hspace{.1pt}\discretionary{.}{%
}{.}\hspace{.4pt}2744098}}}


\bibitem{https://doi.org/10.1111/cgf.14290}
P.~Xie, W.~Tao, J.~Li, W.~Huang, and S.~Chen.
\newblock Exploring multi-dimensional data via subset embedding.
\newblock {\em Computer Graphics Forum}, 40(3):75--86, 2021. \href{https://doi.org/10.1111/cgf.14290}
{doi: {{%
10\hspace{.1pt}\discretionary{.}{%
}{.}\hspace{.4pt}1111\discretionary{/}{%
}{/}cgf\hspace{.1pt}\discretionary{.}{%
}{.}\hspace{.4pt}14290}}}


\bibitem{DBLP:journals/tvcg/XieDW21}
X.~Xie, F.~Du, and Y.~Wu.
\newblock A visual analytics approach for exploratory causal analysis: Exploration, validation, and applications.
\newblock {\em IEEE Transactions on Visualization and Computer Graphics}, 27(2):1448--1458, 2021. \href{https://doi.org/10.1109/TVCG.2020.3028957}
{doi: {{%
10\hspace{.1pt}\discretionary{.}{%
}{.}\hspace{.4pt}1109\discretionary{/}{%
}{/}TVCG\hspace{.1pt}\discretionary{.}{%
}{.}\hspace{.4pt}2020\hspace{.1pt}\discretionary{.}{%
}{.}\hspace{.4pt}3028957}}}


\bibitem{DBLP:journals/tvcg/XiongSHF20}
C.~Xiong, J.~Shapiro, J.~Hullman, and S.~Franconeri.
\newblock Illusion of causality in visualized data.
\newblock {\em IEEE Transactions on Visualization and Computer Graphics}, 26(1):853--862, 2020. \href{https://doi.org/10.1109/TVCG.2019.2934399}
{doi: {{%
10\hspace{.1pt}\discretionary{.}{%
}{.}\hspace{.4pt}1109\discretionary{/}{%
}{/}TVCG\hspace{.1pt}\discretionary{.}{%
}{.}\hspace{.4pt}2019\hspace{.1pt}\discretionary{.}{%
}{.}\hspace{.4pt}2934399}}}


\bibitem{10.1145/3444944}
L.~Yao, Z.~Chu, S.~Li, Y.~Li, J.~Gao, and A.~Zhang.
\newblock A survey on causal inference.
\newblock {\em ACM Transactions on Knowledge Discovery from Data}, 15(5),  article no. 74,  46 pages, may 2021. \href{https://doi.org/10.1145/3444944}
{doi: {{%
10\hspace{.1pt}\discretionary{.}{%
}{.}\hspace{.4pt}1145\discretionary{/}{%
}{/}3444944}}}


\bibitem{DBLP:journals/cgf/YenPF19}
C.~E. Yen, A.~G. Parameswaran, and W.~Fu.
\newblock An exploratory user study of visual causality analysis.
\newblock {\em Computer Graphics Forum}, 38(3):173--184, 2019. \href{https://doi.org/10.1111/cgf.13680}
{doi: {{%
10\hspace{.1pt}\discretionary{.}{%
}{.}\hspace{.4pt}1111\discretionary{/}{%
}{/}cgf\hspace{.1pt}\discretionary{.}{%
}{.}\hspace{.4pt}13680}}}


\bibitem{10.1145/3544548.3581021}
C.-H. Yen, H.~Cheng, Y.~Xia, and Y.~Huang.
\newblock Crowdidea: Blending crowd intelligence and data analytics to empower causal reasoning.
\newblock In {\em Proc. CHI}, CHI '23,  article no. 463,  17 pages, 2023. \href{https://doi.org/10.1145/3544548.3581021}
{doi: {{%
10\hspace{.1pt}\discretionary{.}{%
}{.}\hspace{.4pt}1145\discretionary{/}{%
}{/}3544548\hspace{.1pt}\discretionary{.}{%
}{.}\hspace{.4pt}3581021}}}


\bibitem{9906903}
X.~Zhang, J.~P. Ono, H.~Song, L.~Gou, K.-L. Ma, and L.~Ren.
\newblock Sliceteller: A data slice-driven approach for machine learning model validation.
\newblock {\em IEEE Transactions on Visualization and Computer Graphics}, 29(1):842--852, 2023. \href{https://doi.org/10.1109/TVCG.2022.3209465}
{doi: {{%
10\hspace{.1pt}\discretionary{.}{%
}{.}\hspace{.4pt}1109\discretionary{/}{%
}{/}TVCG\hspace{.1pt}\discretionary{.}{%
}{.}\hspace{.4pt}2022\hspace{.1pt}\discretionary{.}{%
}{.}\hspace{.4pt}3209465}}}


\bibitem{zitzler2001spea2}
E.~Zitzler, M.~Laumanns, and L.~Thiele.
\newblock Spea2: Improving the strength pareto evolutionary algorithm.
\newblock {\em TIK report}, 103, 2001. \href{https://doi.org/10.3929/ethz-a-004284029}
{doi: {{%
10\hspace{.1pt}\discretionary{.}{%
}{.}\hspace{.4pt}3929\discretionary{/}{%
}{/}ethz\discretionary{%
}{-}{-}a\discretionary{%
}{-}{-}004284029}}}


\end{thebibliography}

\appendix 

\section{About Appendices}
Refer to \cref{sec:appendices_inst} for instructions regarding appendices.

\section{Troubleshooting}
\label{appendix:troubleshooting}

\subsection{ifpdf error}

If you receive compilation errors along the lines of \texttt{Package ifpdf Error: Name clash, \textbackslash ifpdf is already defined} then please add a new line \verb|\let\ifpdf\relax| right after the \verb|\documentclass[journal]{vgtc}| call.
Note that your error is due to packages you use that define \verb|\ifpdf| which is obsolete (the result is that \verb|\ifpdf| is defined twice); these packages should be changed to use \verb|ifpdf| package instead.

\subsection{\texttt{pdfendlink} error}

Occasionally (for some \LaTeX\ distributions) this hyper-linked bib\TeX\ style may lead to \textbf{compilation errors} (\texttt{pdfendlink ended up in different nesting level ...}) if a reference entry is broken across two pages (due to a bug in \verb|hyperref|).
In this case, make sure you have the latest version of the \verb|hyperref| package (i.e.\ update your \LaTeX\ installation/packages) or, alternatively, revert back to \verb|\bibliographystyle{abbrv-doi}| (at the expense of removing hyperlinks from the bibliography) and try \verb|\bibliographystyle{abbrv-doi-hyperref}| again after some more editing.

\end{document}